\documentclass[twocolumn,aps,tightenlinefs,superscriptaddress,preprintnumbers]{revtex4-2}

\usepackage{graphicx} 
\usepackage{epstopdf} 
\usepackage{dcolumn} 
\usepackage{xcolor} 
\usepackage[colorlinks,linkcolor=blue,anchorcolor=green,citecolor=blue,urlcolor=blue]{hyperref}
\usepackage[utf8]{inputenc} 
\usepackage[english]{babel} 
\usepackage{blindtext} 
\usepackage{ifthen} 
\usepackage{tabularx}
\usepackage{multirow}
\usepackage{appendix}
\usepackage{amsmath}
\usepackage{lineno}
\usepackage{orcidlink} 

\newcolumntype{L}[1]{>{\raggedright\let\newline\\\arraybackslash\hspace{0pt}}m{#1}}
\newcolumntype{C}[1]{>{\centering\let\newline\\\arraybackslash\hspace{0pt}}m{#1}}
\newcolumntype{R}[1]{>{\raggedleft\let\newline\\\arraybackslash\hspace{0pt}}m{#1}}

\graphicspath{{figures/}} 

\newboolean{articletitles}
\setboolean{articletitles}{true} 

\RequirePackage{xspace}
\RequirePackage{relsize}

\def\belletwo{\mbox{Belle~II}\xspace}

\def\babar{\mbox{\slshape B\kern-0.1em{\smaller A}\kern-0.1em
    B\kern-0.1em{\smaller A\kern-0.2em R}}\xspace}

\mathchardef\PDelta="7101
\mathchardef\PXi="7104
\mathchardef\PLambda="7103
\mathchardef\PSigma="7106
\mathchardef\POmega="710A
\mathchardef\PUpsilon="7107
                 
\def\PB      {\ensuremath{B}\xspace}                 
                 
\def\PD      {\ensuremath{D}\xspace}

\def\PK      {\ensuremath{K}\xspace}

\def\Pi      {\ensuremath{i}\xspace}

\def\kaon  {\ensuremath{\PK}\xspace}
\def\Kbar  {\kern 0.2em\overline{\kern -0.2em \PK}{}\xspace}

\def\Kz    {\ensuremath{\kaon^0}\xspace}
\def\Kzb   {\ensuremath{\Kbar^0}\xspace}
\def\KzKzb {\ensuremath{\Kz \kern -0.16em \Kzb}\xspace}
\def\Kp    {\ensuremath{\kaon^+}\xspace}
\def\Km    {\ensuremath{\kaon^-}\xspace}

\def\KpKm  {\ensuremath{\Kp \kern -0.16em \Km}\xspace}

\def\D       {\ensuremath{\PD}\xspace}
\def\Dbar    {\kern 0.2em\overline{\kern -0.2em \PD}{}\xspace}

\def\Dz      {\ensuremath{\D^0}\xspace}
\def\Dzb     {\ensuremath{\Dbar^0}\xspace}
\def\DzDzb   {\ensuremath{\Dz {\kern -0.16em \Dzb}}\xspace}
\def\Dp      {\ensuremath{\D^+}\xspace}
\def\Dm      {\ensuremath{\D^-}\xspace}

\def\DpDm    {\ensuremath{\Dp {\kern -0.16em \Dm}}\xspace}

\def\Bbar    {\ensuremath{\kern 0.18em\overline{\kern -0.18em \PB}{}}\xspace}

\def\Y#1S{\ensuremath{\PUpsilon{(#1S)}}\xspace}

\def\Lbar {\ensuremath{\kern 0.1em\overline{\kern -0.1em\PLambda}}\xspace}

\def\to                 {\ensuremath{\rightarrow}\xspace}

\newcommand{\tev}{\ensuremath{\mathrm{\,Te\kern -0.1em V}}\xspace}
\newcommand{\gev}{\ensuremath{\mathrm{\,Ge\kern -0.1em V}}\xspace}
\newcommand{\mev}{\ensuremath{\mathrm{\,Me\kern -0.1em V}}\xspace}
\newcommand{\kev}{\ensuremath{\mathrm{\,ke\kern -0.1em V}}\xspace}
\newcommand{\ev}{\ensuremath{\mathrm{\,e\kern -0.1em V}}\xspace}
\newcommand{\gevc}{\ensuremath{{\mathrm{\,Ge\kern -0.1em V\!/}c}}\xspace}
\newcommand{\mevc}{\ensuremath{{\mathrm{\,Me\kern -0.1em V\!/}c}}\xspace}
\newcommand{\gevcc}{\ensuremath{{\mathrm{\,Ge\kern -0.1em V\!/}c^2}}\xspace}
\newcommand{\gevgevcccc}{\ensuremath{{\mathrm{\,Ge\kern -0.1em V^2\!/}c^4}}\xspace}
\newcommand{\mevcc}{\ensuremath{{\mathrm{\,Me\kern -0.1em V\!/}c^2}}\xspace}

\def\gsim{{~\raise.15em\hbox{$>$}\kern-.85em
          \lower.35em\hbox{$\sim$}~}\xspace}
\def\lsim{{~\raise.15em\hbox{$<$}\kern-.85em
          \lower.35em\hbox{$\sim$}~}\xspace}

\def\evtgen   {\mbox{\textsc{EvtGen}}}
\def\pythia    {\mbox{\textsc{Pythia}}}
\def\kkmc     {\mbox{\textsc{KKMC}}}
\def\geant     {\mbox{\textsc{Geant}}}
\def\photos   {\mbox{\textsc{Photos}}\xspace}

\begin{document}



\title{\textbf{First measurements of the branching fractions for the decay modes $\Xi_c^{0} \to \Lambda \eta$ and $\Xi_c^0 \to \Lambda \eta'$ and search for the decay $\Xi_c^{0} \to \Lambda \pi^0$ using Belle and Belle~II data}}

  \author{M.~Abumusabh\,\orcidlink{0009-0004-1031-5425}} 
  \author{I.~Adachi\,\orcidlink{0000-0003-2287-0173}} 
  \author{L.~Aggarwal\,\orcidlink{0000-0002-0909-7537}} 
  \author{H.~Ahmed\,\orcidlink{0000-0003-3976-7498}} 
  \author{Y.~Ahn\,\orcidlink{0000-0001-6820-0576}} 
  \author{H.~Aihara\,\orcidlink{0000-0002-1907-5964}} 
  \author{N.~Akopov\,\orcidlink{0000-0002-4425-2096}} 
  \author{S.~Alghamdi\,\orcidlink{0000-0001-7609-112X}} 
  \author{M.~Alhakami\,\orcidlink{0000-0002-2234-8628}} 
  \author{A.~Aloisio\,\orcidlink{0000-0002-3883-6693}} 
% ...
\collaboration{The Belle and Belle II Collaborations}

\begin{abstract}
Using data samples of 988.4~fb$^{-1}$ and 427.9~fb$^{-1}$ collected with the Belle and \belletwo detectors, we present a study of the singly Cabibbo-suppressed decays $\Xi_c^{0} \to \Lambda \eta$, $\Lambda \eta'$, and $\Lambda \pi^0$. We observe the decay $\Xi_c^0 \to \Lambda \eta$ and find evidence for the decay $\Xi_c^0 \to \Lambda \eta'$, with corresponding branching ratios determined to be ${\mathcal{B}(\Xi_c^0 \to \Lambda \eta)}/{\mathcal{B}(\Xi_c^0 \to \Xi^- \pi^+)}= (4.16 \pm 0.91 \pm {0.23})\%$ and ${\mathcal{B}(\Xi_c^0 \to \Lambda \eta')}/{\mathcal{B}(\Xi_c^0 \to \Xi^- \pi^+)}= (2.48 \pm 0.82 \pm {0.12})\%$, respectively. We find no significant signal in the $\Xi_c^0 \to \Lambda \pi^0$ decay mode and set an upper limit at the 90\% credibility level of ${\mathcal{B}(\Xi_c^0 \to \Lambda \pi^0)}/{\mathcal{B}(\Xi_c^0 \to \Xi^- \pi^+)}< {3.5\%}$. Multiplying these ratios by the world-average branching fraction of the normalization channel, $\mathcal{B}(\Xi_c^0 \to \Xi^- \pi^+)=(1.43 \pm 0.27)\%$, we obtain the absolute branching fractions of $\mathcal{B}(\Xi_c^0 \to \Lambda \eta)= (5.95 \pm 1.30 \pm {0.32} \pm 1.13) \times 10^{-4}$, $\mathcal{B}(\Xi_c^0 \to \Lambda \eta')= (3.55 \pm 1.17 \pm {0.17} \pm 0.68) \times 10^{-4}$, and an upper limit at the 90\% credibility level on the absolute branching fraction of $\mathcal{B}(\Xi_c^0 \to \Lambda \pi^0)< {5.2} \times 10^{-4}$. The quoted first and second uncertainties are statistical and systematic, respectively, while the third uncertainties arise from the branching fraction of the normalization mode. These results are consistent with most theoretical predictions and further the understanding of the underlying decay mechanisms.
\end{abstract}

\keywords{Charmed baryon, Singly Cabibbo-suppressed decay, $e^+e^-$ experiments, Belle~(II)}

\maketitle

\section{Introduction}

Charmed baryon spectroscopy serves as an important platform for investigating the dynamics of light quarks in the presence of one or two heavy quarks. Topological diagrams are used extensively to explain the mechanisms in nonleptonic weak decays of charmed baryons~\cite{Chau:1995gk,Cheng:2021qpd}. According to the topological diagrams, both factorizable and non-factorizable contributions are significant in baryon decay amplitudes~\cite{Cheng:2018hwl,Cheng:2021qpd}. Precise measurements of the branching fractions will help to clarify the theoretical picture and provide a clearer understanding of the decay dynamics. 

Over the past decade, significant advancements have been made in the experimental study of charmed baryon physics~\cite{ParticleDataGroup:2024cfk}. {The antitriplet charmed baryons, characterized by their purely weak decays, serve as foundational systems in this field, anchoring both high-precision measurements and theoretical frameworks based on SU(3) flavor symmetry.} The absolute branching fractions of three key decays $\Lambda_c^+ \to p K^- \pi^+$~\cite{Belle:2013jfq,BESIII:2015bjk}, $\Xi_c^+ \to \Xi^- \pi^+ \pi^+$~\cite{Belle:2019bgi}, and $\Xi_c^0 \to \Xi^- \pi^+$~\cite{Belle:2018kzz} have been measured. Charge-conjugate modes are included throughout the analysis. These results provide a foundation for determining {absolute} branching fractions in antitriplet charm baryon decays~\cite{LHCb:2017xtf,Belle:2021mvw,LHCb:2019nxp,Belle:2024xcs, Belle:2021avh,Belle:2024ikp}. {Comprehensive and precise experimental measurements are essential for testing different theoretical models and for clarifying the decay mechanisms of antitriplet charmed baryons. Theoretical calculations of the two-body hadronic weak decays of $\Xi_c^0$ have been performed with topological diagrams~\cite{Zhao:2018mov,Hsiao:2021nsc,Zhong:2024zme}.} However, many branching fractions remain unmeasured, particularly for Cabibbo-suppressed decay modes.

\begin{figure*}[!htbp]
    \centering
    \includegraphics[width=1.0\linewidth]{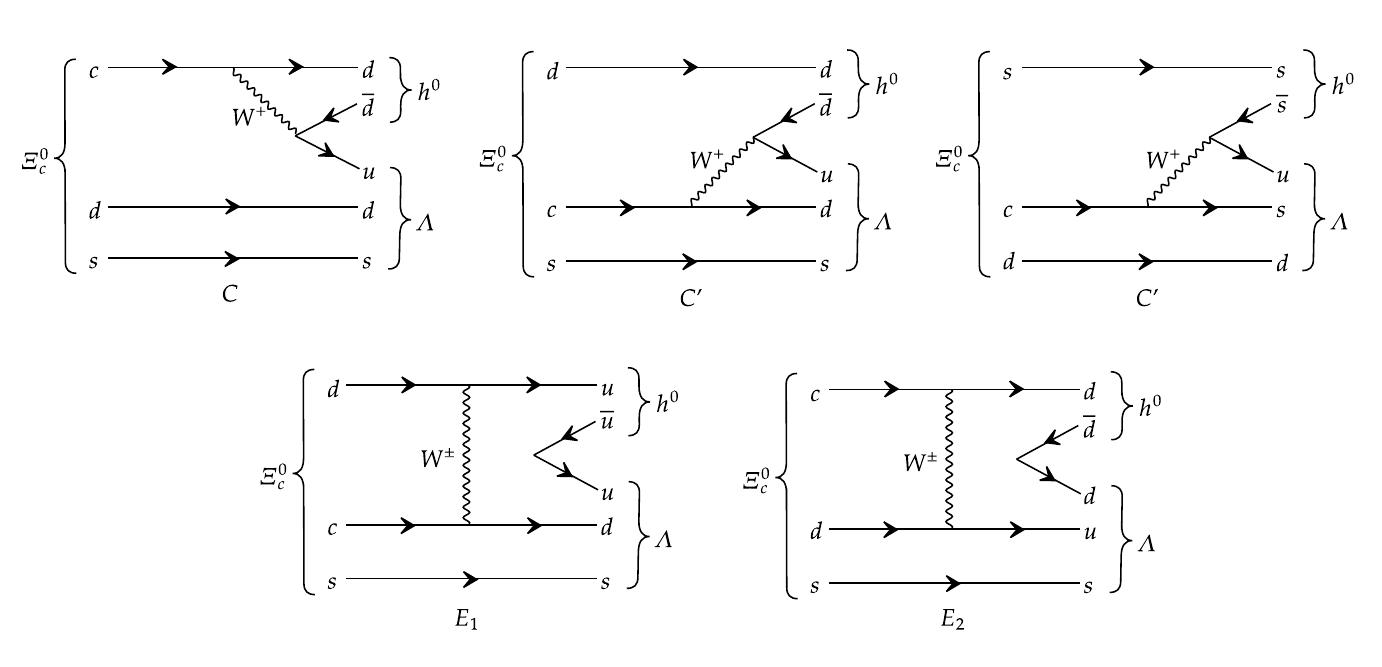}
    \caption{Topological diagrams contributing to $\Xi_c^0 \to \Lambda h^0$ decays: internal $W$-emission $C$, inner $W$-emission $C'$, {and} $W$-exchange diagrams $E_1$ and $E_2$~\cite{Cheng:2021qpd}.}
    \label{fig:topo_diag}
\end{figure*}

In this paper, we present the first measurements of the branching fractions of the singly Cabibbo-suppressed decays $\Xi_c^0 \to \Lambda h^0$ $(h=\eta,~\eta',~\pi^0)$, {relative} to the decay $\Xi_c^0 \to \Xi^- \pi^+$. The {topological diagrams} for the signal decays are illustrated in Fig.~\ref{fig:topo_diag}. {Decay amplitudes {have} two main components: (i) factorizable contributions from internal $W$-emission processes ($C$), and (ii) non-factorizable contributions that arise from inner $W$-emission ($C'$) and $W$-exchange mechanisms ($E_1$ and $E_2$)~\cite{Chau:1995gk,Cheng:2021qpd}.} {Theoretical predictions for these branching fractions are primarily derived from an analysis of these diagrams based on SU(3) flavor symmetry~\cite{Zhao:2018mov,Hsiao:2021nsc,Zhong:2024zme}, irreducible SU(3)~\cite{Geng:2018plk,Geng:2019xbo,Zhong:2022exp,Xing:2024nvg}, or a pole model~\cite{Zou:2019kzq}. Across these different frameworks, the predicted values span a broad range from $10^{-5}$ to $10^{-3}$~\cite{Zhao:2018mov,Geng:2018plk,Zou:2019kzq,Geng:2019xbo,Hsiao:2021nsc,Zhong:2022exp,Geng:2023pkr,Zhong:2024zme,Xing:2024nvg}. A representative example of the resulting theoretical spread is provided for $\Xi_c^0 \to \Lambda \eta'$, whose predictions vary from $(0.2\pm0.1) \times 10^{-4}$~\cite{Zhao:2018mov} to $(16.4\pm10.6) \times 10^{-4}$~\cite{Geng:2019xbo}, depending on the specific model scenario adopted.}

\section{The Belle and Belle~II detectors}

The Belle detector~\cite{Belle:2000cnh}, a large solid-angle magnetic spectrometer, operated at the KEKB asymmetric-energy $e^+e^-$ collider~\cite{Kurokawa:2001nw,Abe:2013kxa} from 1999 to 2010. It comprised a silicon vertex detector~(SVD), a 50-layer central drift chamber~(CDC), an array of aerogel threshold Cherenkov counters~(ACC), a barrel-like arrangement of time-of-flight scintillation counters~(TOF), and an electromagnetic calorimeter~(ECL) with 8736 CsI(Tl) crystals. These components were all located inside a superconducting solenoid providing a magnetic field of 1.5~T. An iron flux-return yoke placed outside the solenoid coil was instrumented with resistive plate chambers to detect $K^0_{L}$ mesons and muons~(KLM). Further details are provided in Ref.~\cite{Belle:2000cnh}.

The Belle~II detector~\cite{Abe:2010gxa}, located at the SuperKEKB asymmetric-energy $e^+e^-$ accelerator~\cite{Akai:2018mbz}, collects data at and near the $\Upsilon(4S)$ resonance since 2019. {The Belle~II detector is based on the Belle detector but contains several new subsystems as well as substantial upgrades to others.} The innermost subdetector is the vertex detector (VXD), which consists of a two-layer silicon-pixel detector~(PXD) surrounded by a four-layer SVD~\cite{Belle-IISVD:2022upf}. Only one sixth of the second layer of the PXD was installed for the data analyzed here. {Together with the VXD, a new large-radius, helium-ethane, small-cell CDC reconstructs tracks of charged particles.} Surrounding the CDC, which also provides energy-loss measurements, is a time-of-propagation counter~(TOP)~\cite{Atmacan:2025jmh} in the central region and an aerogel-based ring-imaging Cherenkov counter~(ARICH) in the forward region. These detectors provide
charged-particle identification. {The Belle CsI(Tl) crystal ECL, the Belle solenoid and the iron flux-return yoke are reused in the Belle~II detector. The ECL readout electronics have been upgraded and the instrumentation in the flux-return yoke to identify $K_L^0$ mesons and muons has been replaced.} The $z$-axis is defined as the central solenoid axis, with the positive direction oriented toward the electron beam, for both Belle and Belle~II.

\section{Data samples}

This analysis uses an integrated luminosity of 1.42~$\text{ab}^{-1}$, comprising 988.4~$\text{fb}^{-1}$~\cite{Brodzicka:2012jm} from the Belle detector~\cite{Belle:2000cnh} and 427.9~$\text{fb}^{-1}$~\cite{Belle-II:2024vuc} from the Belle~II detector. The Belle detector collected data at or near the $\Upsilon(nS)$ ($n=$1, 2, 3, 4, 5) resonances, whereas the Belle~II detector accumulated them at or near the $\Upsilon(4S)$ and at or near 10.75~GeV in the center-of-mass (c.m.)\ frame.

Monte Carlo~(MC) samples generated with \evtgen~\cite{Lange:2001uf} are used to optimize signal selection criteria and calculate selection efficiencies. Continuum $e^+e^- \to c\bar{c}$ events are generated using \pythia6~\cite{Sjostrand:2000wi} for Belle {and \kkmc~\cite{Jadach:1999vf} interfaced to \pythia8~\cite{Sjostrand:2014zea}} for Belle~II. In signal MC events, a charm quark hadronizes into a $\Xi_c^0$, with decays $\Xi_c^0 \to \Lambda \eta,~\Lambda \eta'~\text{or}~\Lambda \pi^0$ generated by a {phase-space model (when summing over charge-conjugate states possible polarization effects cancel~\cite{CLEO:2000lsg}).} The final-state radiation is simulated via {the} \photos package~\cite{Barberio:1990ms}. The Belle and Belle~II {detector responses} are simulated using \geant3~\cite{Brun:1987ma} and \geant4~\cite{Agostinelli:2002hh}, respectively. The inclusive MC samples are four times the integrated luminosity of the Belle and Belle~II data, including $\Upsilon(1S,~2S,~3S)$ decays, $\Upsilon(4S) \to B{\bar B}$, $\Upsilon(5S) \to B_{(s)}^{(*)}{\bar B}_{(s)}^{(*)}$, and $e^+ e^- \to q{\bar q}~(q = u,~d,~s,~c)$~\cite{Zhou:2020ksj}. 

\section{Event selection criteria}

We reconstruct the decays $\Xi_c^0 \to \Lambda \eta,~\Lambda \eta'~\text{and}~ \Lambda \pi^0$, with subsequent decays $\Lambda \to p \pi^-$, $\eta' \to \pi^+ \pi^- \eta$, $\eta' \to \pi^+ \pi^- \gamma$, $\eta \to \gamma \gamma$, $\eta \to \pi^+ \pi^- \pi^0$, and $\pi^0 \to \gamma \gamma$. {All Belle data and MC samples} are converted to the Belle~II format using B2BII software package~\cite{Gelb:2018agf}. This allows for unified analysis of both Belle and Belle~II samples within the Belle~II analysis software framework ({basf2})~\cite{Kuhr:2018lps}. The selection criteria are optimized to enhance the sensitivity to signal candidates by maximizing the Punzi figure of merit (FOM), defined as FOM $=\epsilon_{\Lambda h^0}/(3/2+\sqrt{N_\mathrm{b}})$~\cite{Punzi:2003bu,Punzi:2003wze,Feichtinger:2021uff}. Here, $\epsilon_{\Lambda h^0}$ is the selection efficiency from signal MC samples, and $N_\mathrm{b}$ is the background yield obtained from inclusive MC samples in the respective $\Xi_c^0$ signal regions: (i) $|M(\Lambda\eta)-m_{\Xi_c^0}|<38/20$~MeV/$c^2$ for the two $\eta$ decay channels $\eta \to \gamma \gamma$ and $\eta \to \pi^+ \pi^- \pi^0$; (ii) $|M(\Lambda\eta')-m_{\Xi_c^0}|<16/11/11$~MeV/$c^2$ for the three $\eta'$ decay channels $\eta' \to \pi^+ \pi^- \eta$ ($\eta \to \gamma \gamma$), $\eta' \to \pi^+ \pi^- \eta$ ($\eta \to \pi^+ \pi^- \pi^0$), and $\eta' \to \pi^+ \pi^- \gamma$; and (iii) $-$83~MeV/$c^2<M(\Lambda\pi^0)-m_{\Xi_c^0}<55$~MeV/$c^2$ {because the invariant mass distribution exhibits asymmetric tails}. Each signal region corresponds to approximately 3 standard deviations (3$\sigma$) {of the invariant mass peak of the respective decay channel}. Here and throughout this paper, $M({\rm AB})$ stands for the invariant mass of the system of A and B particles, and $m_{X}$ represents the nominal mass of particle $X$~\cite{ParticleDataGroup:2024cfk}. We apply identical selection criteria to the Belle and Belle~II analyses unless otherwise specified.

Charged {particles} {not used for $\Lambda$ reconstruction} are required to have impact parameters relative to the $e^+e^-$ interaction point (IP) of less than 0.5~cm perpendicular to the $z$-axis and less than 2.0~cm parallel to it. For the particle identification~(PID), information from different detector subsystems of Belle or Belle~II is combined to form binary likelihood {ratios}, $\mathcal{R}(i|j)=\mathcal{L}_{i}/(\mathcal{L}_{i}+\mathcal{L}_{j})$, where $\mathcal{L}_{i}$  and $\mathcal{L}_{j}$ represent the likelihoods of the track being identified as hadrons $\pi$, $K$, or $p$, as appropriate. Tracks with $\mathcal{R}(p|K)>0.6$ and $\mathcal{R}(p|\pi)>0.6$ are identified as protons. {The other tracks with $\mathcal{R}(\pi|K)>0.6$ are identified as pions.} The resulting PID efficiencies range from 85\% to 96\%, with corresponding misidentification rate between 1\% and 5\%.

Neutral clusters are used as photon candidates if they are not matched to any charged track. To suppress background from neutral hadrons, {we reject photon candidates if the ratio of energies deposited in the central $3\times3$ square of cells to that deposited in the enclosing $5\times5$ square of cells (with corner cells omitted in Belle~II) in its ECL cluster is less than 0.85. The exclusion of corners in Belle~II mitigates the increased background contribution due to its higher instantaneous luminosity.} In the c.m.\ frame, the photon-energy ($E_{\gamma}^{\rm c.m.}$) threshold is 280~MeV for $\eta\to\gamma\gamma$ in $\Xi_{c}^{0}\to\Lambda\eta$, 100~MeV {for $\eta' \to \eta \pi^+ \pi^-, \eta \to \gamma \gamma$}, 320~MeV for $\eta'\to\pi^{+}\pi^{-}\gamma$, and 70~MeV for $\pi^{0}$ reconstruction. {We use the FastBDT classifier~\cite{Keck:2017gsv,Cheema:2024iek} for both fake-photon rejection and beam-background-cluster rejection, and the average signal efficiencies (background-rejection efficiencies) of them are approximately 95\% (33\%) and 98\% (11\%), respectively. In the FastBDT trainings, the Belle detector provides less information than the Belle~II detector, especially for low-energy photons, which results in a lower efficiency for the decay mode $\Xi_c^0 \to \Lambda \pi^0$ in Belle than in Belle~II.} Detailed descriptions of the classifiers for separating signal photons from fake photons and beam background clusters are provided in Ref.~\cite{Cheema:2024iek}. However, we {do not use} the two FastBDT requirements in the Belle analysis of the decay channels {with $\eta \to \pi^+ \pi^- \pi^0$ decay mode}, as they provide negligible improvement in signal-background separation while {notably} reducing signal efficiency.

{To suppress background {from misreconstructed} $\pi^0$ candidates, a $\pi^0$ veto is applied to the photons used in the reconstruction of the signal $\Xi_c^0$.} The invariant mass of the photon pair, $M(\gamma_{\rm sig} \gamma_{\rm roe})$, must satisfy $|M(\gamma_{\rm sig} \gamma_{\rm roe})-m_{\pi^0}|>9$~MeV/$c^2$ {with a signal efficiency of 80\%} for the $\Xi_c^0 \to \Lambda \eta, \eta \to \gamma \gamma$ decay mode and $|M(\gamma_{\rm sig} \gamma_{\rm roe})-m_{\pi^0}|>13$~MeV/$c^2$ {with a signal efficiency of 85\%} for the $\Xi_c^0 \to \Lambda \eta', \eta' \to \gamma \pi^+ \pi^-$ decay mode. Here, $\gamma_{\rm sig}$ indicates photons {used for signal reconstruction}, whereas $\gamma_{\rm roe}$ denotes photons from the rest of {the event} that are unrelated to the signal candidates.

The $\eta$ meson is reconstructed from both $\gamma \gamma$ and $\pi^+ \pi^- \pi^0$ decays with signal regions defined as $|M(\gamma\gamma)-m_{\eta}|<22$~MeV/$c^2$, {yielding a signal efficiency of 88\%} and $|M(\pi^+\pi^-\pi^0)-m_{\eta}|<6$~MeV/$c^2$, {corresponding to an 85\% signal efficiency}. {The $\eta'$ meson is reconstructed via three decay channels: (i) $\pi^+ \pi^- \eta$ with the $\eta$ further decaying into $\gamma\gamma$, requiring $|M(\pi^+\pi^-\eta)-m_{\eta'}|<6$~MeV/$c^2$ with a signal efficiency of 90\%, (ii) $\pi^+ \pi^- \eta$ with the $\eta$ reconstructed in $\pi^+ \pi^- \pi^0$ and $|M(\pi^+\pi^-\eta)-m_{\eta'}|<9$~MeV/$c^2$ with a signal efficiency of 85\%, and (iii) $\pi^+ \pi^- \gamma$ with the signal region $|M(\pi^+\pi^-\gamma)-m_{\eta'}|<11$~MeV/$c^2$, yielding an 80\% signal efficiency.} The signal $\pi^0$ candidate is formed through its decay into two photons with the signal region $|M(\gamma\gamma)-m_{\pi^0}|<18$~MeV/$c^2$ {with a signal efficiency of 98\%}. Additionally, all $h^0$ candidates are required to have momenta greater than 1.10~GeV/$c$. 

The $\Lambda$ candidates are reconstructed {with} the decay $\Lambda \to p \pi^{-}$. In the Belle analysis, the {$\Lambda$} candidates are selected based on {five} parameters: the distance of two daughter tracks at their interception position in $z$-axis; the minimum distance between daughter tracks and the IP in the transverse {(perpendicular to the $z$-axis)} plane; the angular difference between the $\Lambda$ flight direction and the direction between the IP and the $\Lambda$ decay vertex in the transverse plane; and the flight length of $\Lambda$ {candidates in the transverse} plane. In the Belle~II analysis, the $\Lambda$ candidates {must satisfy} the criteria on three parameters: the cosine of the angle between {the $\Lambda$} momentum and vertex vector (vector connecting IP and {$\Lambda$ decay} vertex); the significance of the $\Lambda$'s flight distance {(the distance in units of its uncertainty)}; the ratio of the proton's momentum to the $\Lambda$'s momentum. The signal region of the reconstructed $\Lambda$ candidates is defined as $|M(p\pi^-)-m_{\Lambda}|<3$~MeV/$c^2$($\sim$$3\sigma$). Additionally, the momentum of $\Lambda$ candidates in the c.m.\ frame must exceed 1.3/1.1/1.5 GeV/$c$ for the $\Lambda \eta/\Lambda \eta'/\Lambda \pi^0$ channels, respectively.

The $\Xi_c^0$ candidates are reconstructed from the $\Lambda \eta$, $\Lambda \eta'$, and $\Lambda \pi^0$ combinations. A TreeFit algorithm~\cite{Krohn:2019dlq} is employed to ensure that decay daughters share a common vertex of origin, with constraints also applied to the masses of intermediate states. {To suppress combinatorial backgrounds}, we require the scaled momentum $x_{p}={p^{\rm c.m.}_{\Xi_c^0}c/\sqrt{s/4-m_{\Xi_c^{0}}^2 c^{4}}}$ to be greater than 0.60 in both Belle and Belle~II analyses, where $p_{\Xi_c^0}^{\rm c.m.}$ is the momentum of $\Xi_c^{0}$ in the c.m.\ frame of $e^+e^{-}$ system, $\sqrt{s}$ is the collision energy, and $m_{\Xi_c^{0}}$ is the invariant mass of {the} $\Xi_c^0$ particle. {Since the momenta of $\Xi_c^0$ candidates produced in continuum processes are higher than those from $B$-meson decays, this requirement also removes the events from $B$ decays.}

To determine the {signal yield} and the signal efficiency of the {normalization} mode, $\Xi_c^0 \to \Xi^- \pi^+$, we apply the selection criteria detailed in Ref.~\cite{Belle:2024ikp} except that we require $x_{p} > 0.60$.

\section{Measurements of the branching fractions}

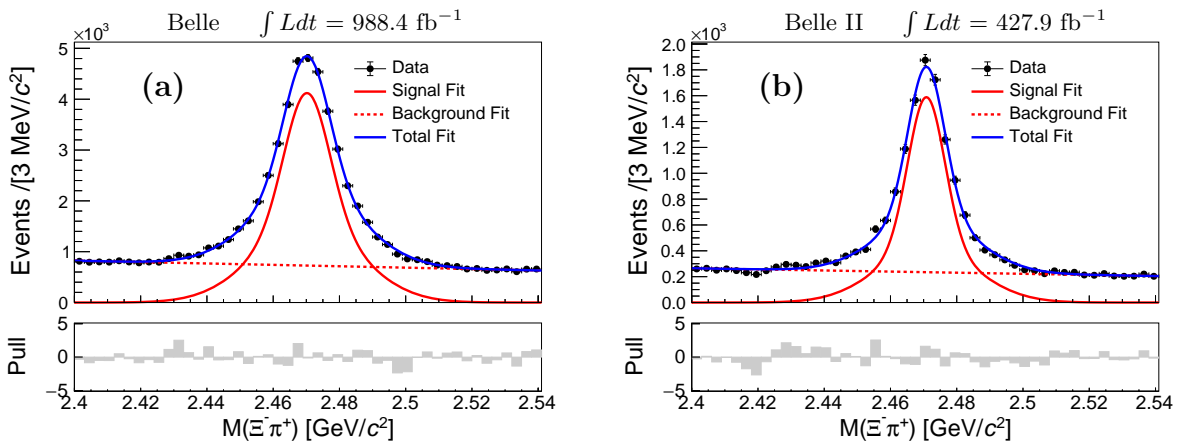
\begin{figure*}
    \begin{tabular}{c c}
        \includegraphics[width=0.45\linewidth]{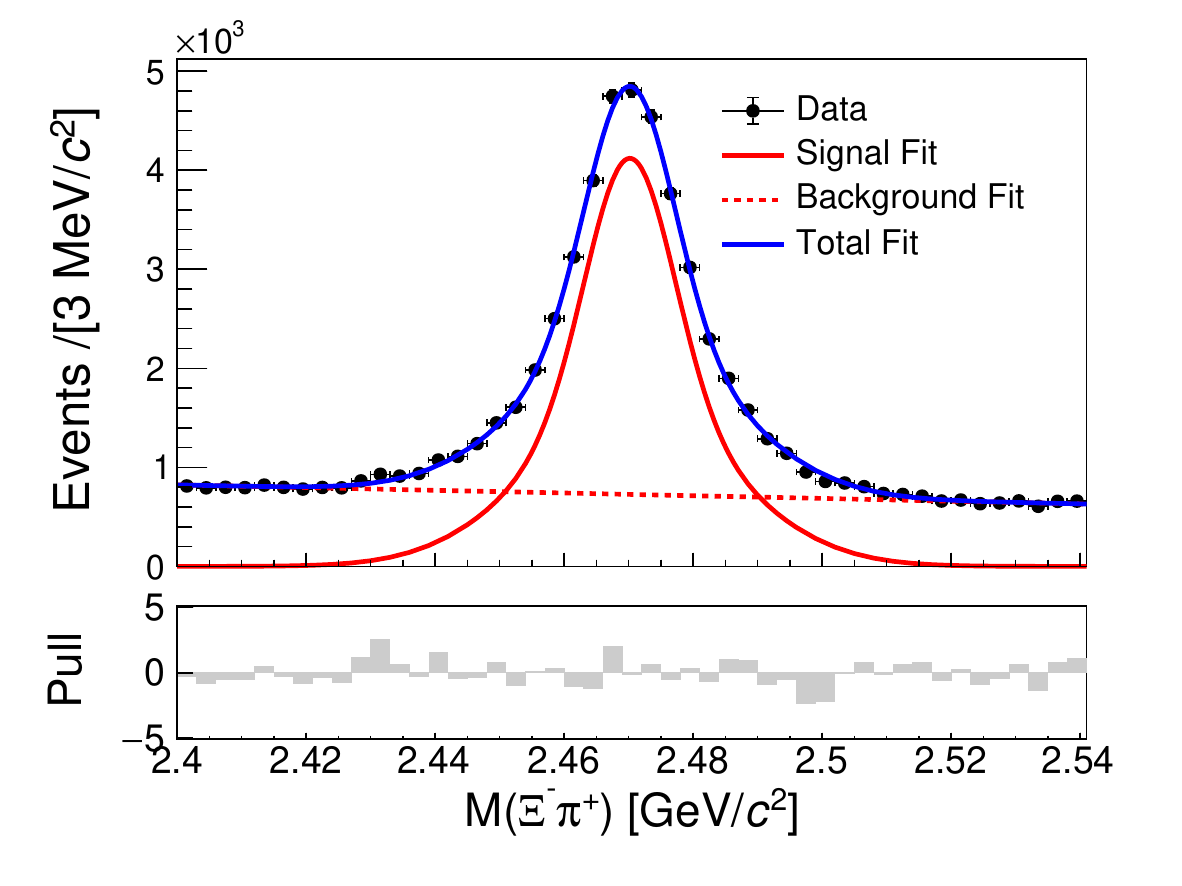}\put(-170,140){\bf \large (a)}
        \put(-160,165){Belle \hspace{0.3cm} $\int L dt$ = 988.4~fb$^{-1}$}
        \put(-185,120){\textcolor{gray}}&
        \includegraphics[width=0.45\linewidth]{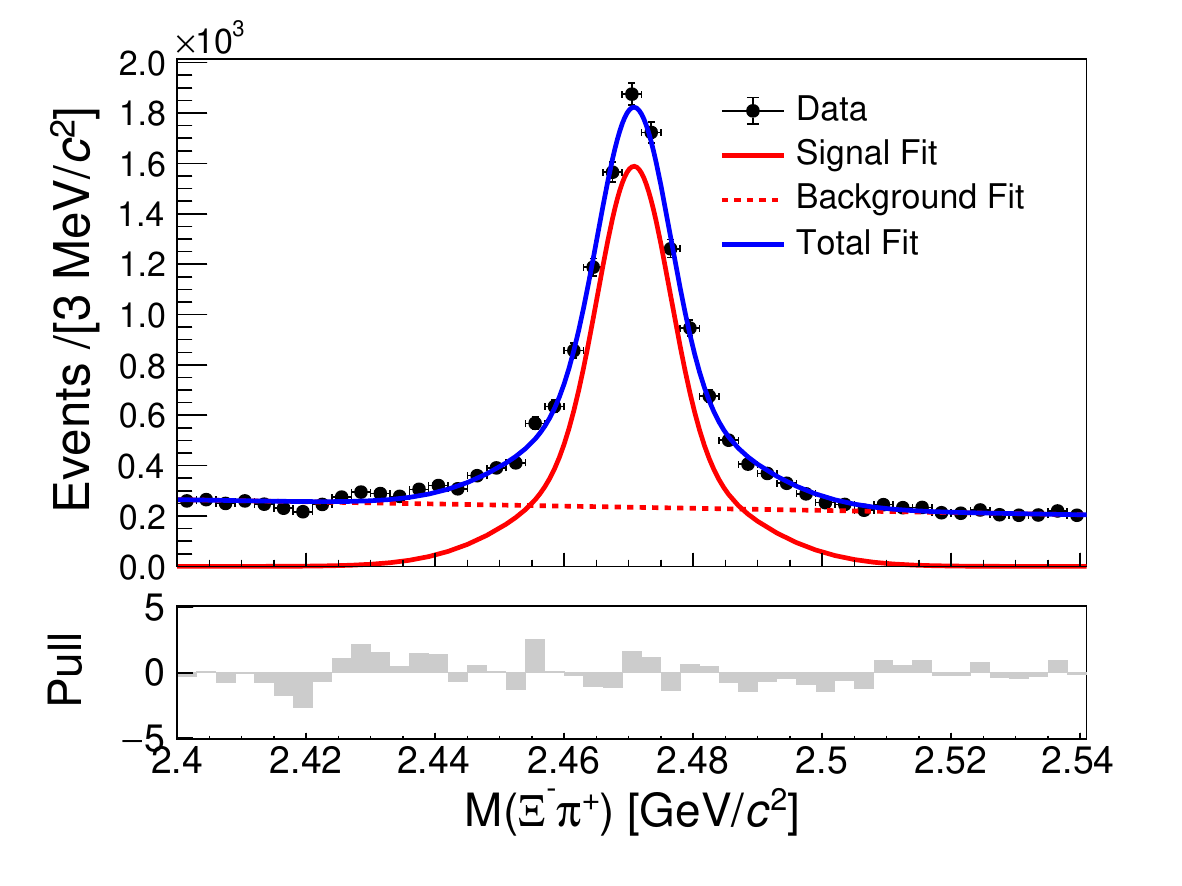}\put(-170,140){\bf \large (b)}
        \put(-160,165){Belle~II \hspace{0.3cm} $\int L dt$ = 427.9~fb$^{-1}$}
        \put(-185,120){\textcolor{gray}}\\
    \end{tabular}
    \caption{The invariant mass spectra of $\Xi^- \pi^+$ candidates in (a) Belle and (b) Belle~II data. Dots with error bars represent the data; solid red curves indicate the signal fit {functions}; dashed red lines denote the fitted combinatorial backgrounds; solid blue curves illustrate the best fit results. {The gray bars show the pull distributions of the best fit results.}}\label{fig:InvM_ref}
\end{figure*}

{For the normalization mode $\Xi_c^0 \to \Xi^- \pi^+$,} signal {probability density functions (PDFs)} are modeled using the double-Gaussian functions, while combinatorial backgrounds are described by second-order Chebyshev polynomials. {All parameters of the signal and combinatorial background shapes are free in the fit.} The {unbinned extended maximum-likelihood} fits to the $\Xi^- \pi^+$ invariant mass distributions are presented in Fig.~\ref{fig:InvM_ref}. For Belle, the signal yield and efficiency are $30230 \pm 281$ and $(11.76 \pm 0.05)\%$, respectively. For Belle~II, they are $11579 \pm 161$ and $(11.35 \pm 0.03)$\%.

\begin{figure*}[!ht]
    \centering
    \begin{tabular}{c c}
        \includegraphics[width=0.45\textwidth]{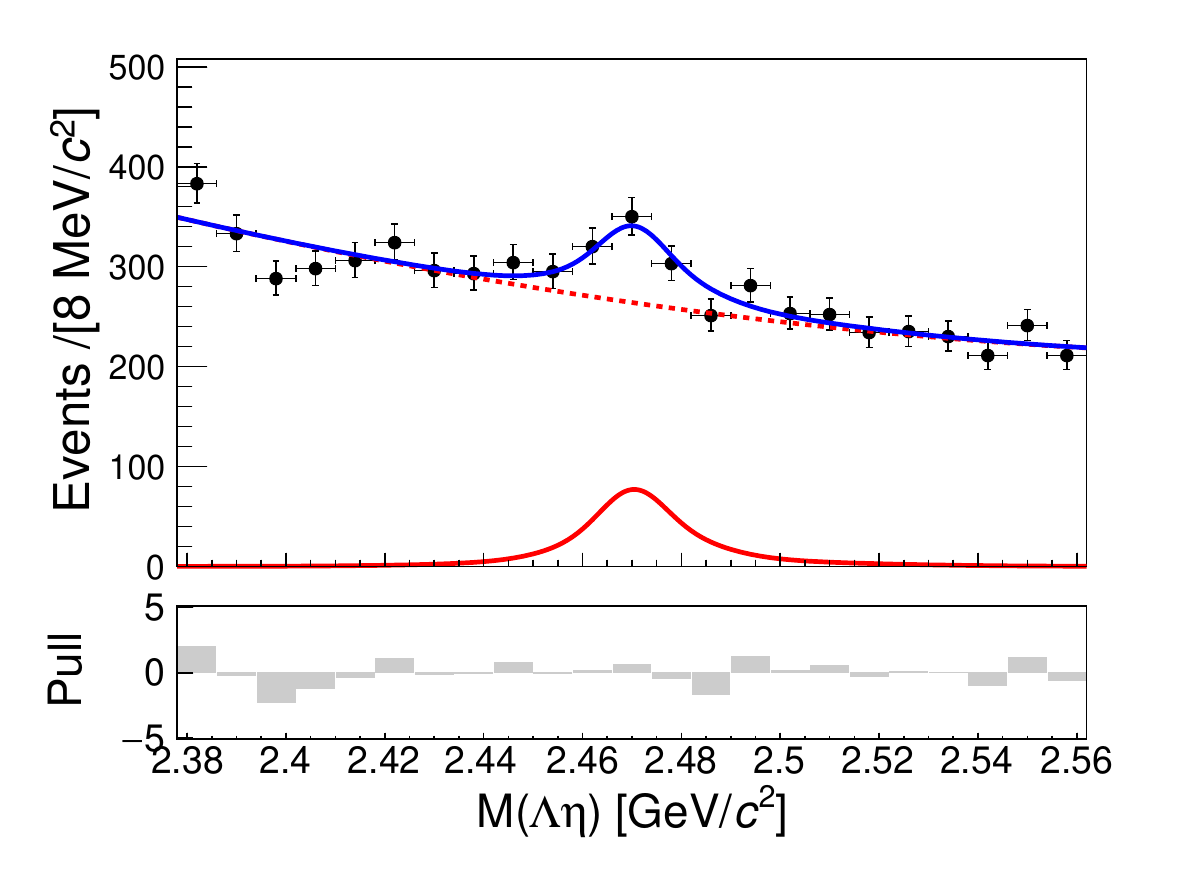}\put(-60,140){\bf \large (a)}
        \put(-180,145){\textcolor{gray}}&
        \includegraphics[width=0.45\textwidth]{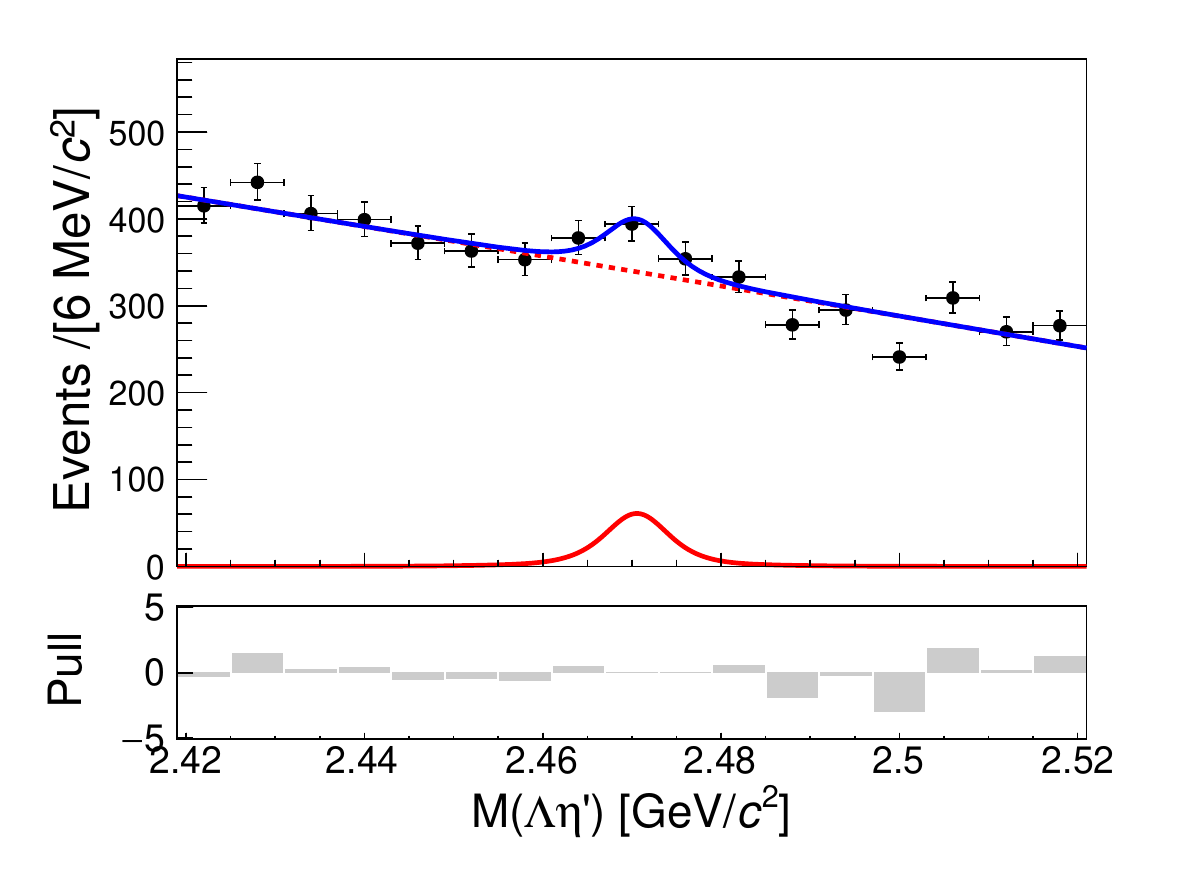}\put(-60,140){\bf \large (b)}
        \put(-180,145){\textcolor{gray}}\\
        \includegraphics[width=0.45\textwidth]{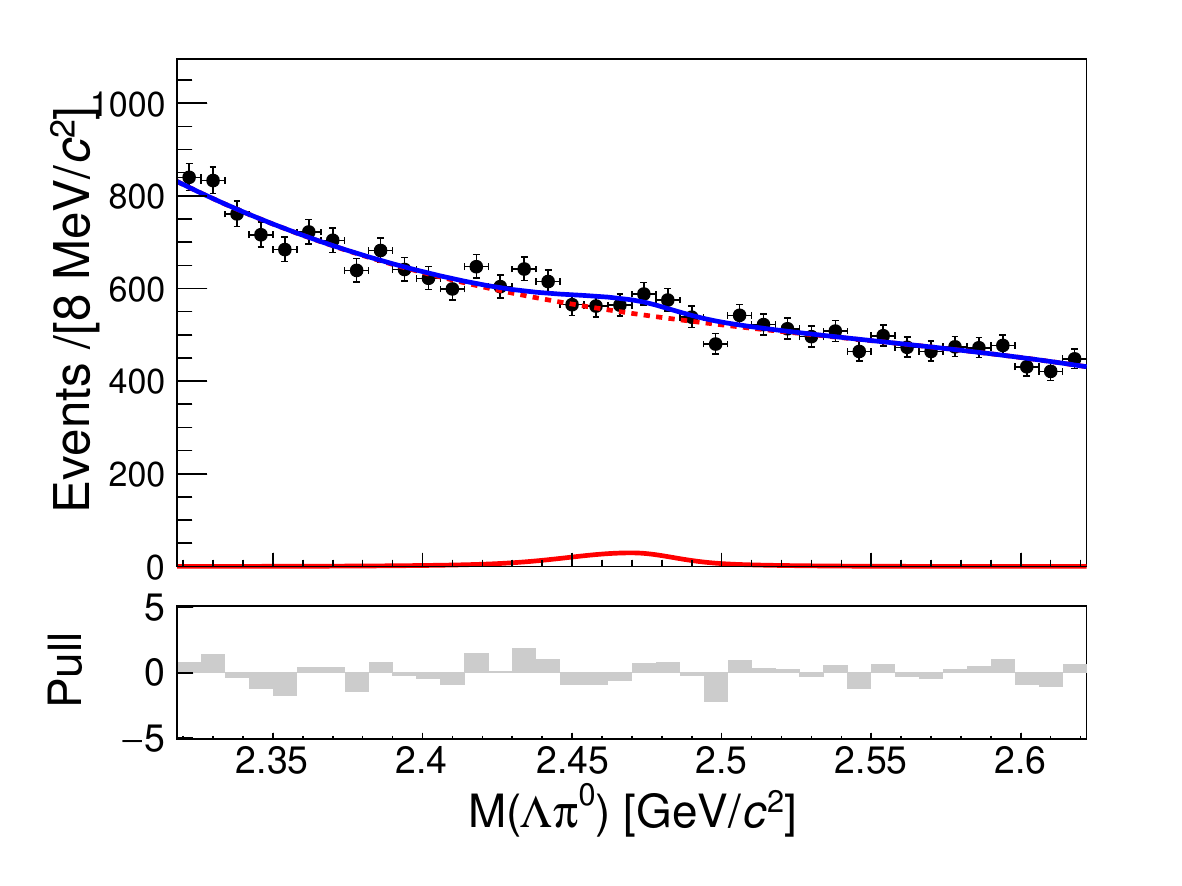}\put(-60,140){\bf \large (c)}
        \put(-180,145){\textcolor{gray}}&
        \begin{minipage}[t]{0.45\textwidth}
            \vspace{-17em}
            \centering
            \large Belle \hspace{0.75cm} $\int L dt = 988.4$ fb$^{-1}$\\[2mm]
            \large Belle~II \hspace{0.3cm} $\int L dt = 427.9$ fb$^{-1}$\\
            \includegraphics[width=\textwidth]{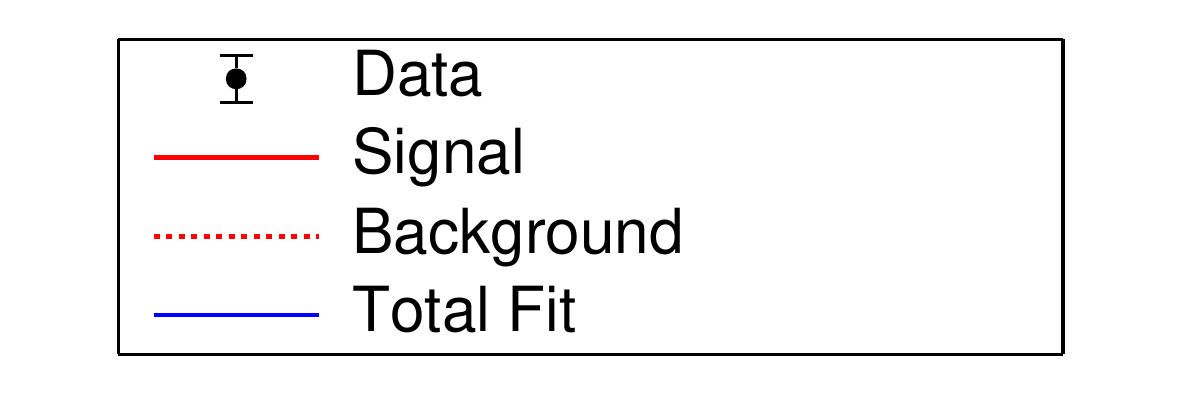}
        \end{minipage}\\
    \end{tabular}
    \caption{The invariant mass spectra of (a) $\Lambda \eta$, (b) $\Lambda \eta'$, and (c) $\Lambda \pi^0$ {candidates overlaid with the fit results {obtained using} Belle and Belle~II data samples.} Dots with error bars represent the {number of events in data}; solid red curves indicate the signal {PDFs}; dashed red lines denote the fitted combinatorial backgrounds; solid blue curves illustrate the fit results. {The gray bars show the pull distributions of the fit results.}}\label{fig:fit_Br}
\end{figure*}

To extract the signal yields for $\Xi_c^0 \to \Lambda \eta$, $\Lambda \eta'$, and $\Lambda \pi^0$ in data, we {conduct} unbinned extended maximum-likelihood fits to the invariant mass distributions $M(\Lambda \eta)$, $M(\Lambda \eta')$, and $M(\Lambda \pi^0)$. The signal shapes for $\Xi_c^0$ candidates reconstructed with $\Lambda \eta$ and $\Lambda \eta'$ modes are modeled using double-Gaussian functions with the same mean values, while the signal shape from $\Lambda \pi^0$ mode is described by two bifurcated Gaussians with distinct mean values. All parameters of signal shapes are fixed to the values obtained from the corresponding signal simulations. {According to the MC simulation, no peaking background component is expected~\cite{Zhou:2020ksj}.} Thus, the combinatorial backgrounds are parametrized using a second-order Chebyshev polynomial whose parameters are left free. Simultaneous fits are performed across different $h^0$ decay modes and datasets from Belle and Belle~II, weighted by the product of {integrated} luminosity $L$, signal efficiency $\epsilon_{\Lambda h^0}$, and product branching fractions $\mathcal{B}(h^0)$ for each mode. {Taking $\Xi_c^0 \to \Lambda \eta$ as an example, the expected signal yield $N_{i,\Lambda \eta}^{\rm exp}$ is proportional to $L^j \cdot \epsilon_{i,\Lambda \eta}^{j} \cdot \mathcal{B}_{i} (\eta)$. Here, the subscript $i$ denotes the $\eta$ decay mode ($\eta \to \gamma \gamma$ or $\eta \to \pi^+ \pi^- \pi^0$), and the superscript $j$ identifies the dataset (Belle or Belle~II) from which the quantities are derived.} Figure~\ref{fig:fit_Br} overlays the combined Belle and Belle~II data, showing simultaneous fits together with the pull distributions, {which are defined as ($N_{\rm data}-N_{\rm fit})/\sqrt{N_{\rm data}}$.} The fitted signal yields are summarized in Table~\ref{tab:eff_Br}. {Fits to the data return signal yields of $262 \pm 57$, $101 \pm 33$, and $190 \pm 120$ events for the $\Xi_c^0 \to \Lambda \eta$, $\Xi_c^0 \to \Lambda \eta'$, and $\Xi_c^0 \to \Lambda \pi^0$ decay modes, corresponding to statistical significances of 5.3$\sigma$, 3.3$\sigma$, and 1.4$\sigma$, respectively. {After including systematic uncertainties (discussed in Section~\ref{sec:uncer}), the $\Lambda \eta$ and $\Lambda \eta'$ modes have signal significances of 5.1$\sigma$ and 3.2$\sigma$, respectively.} These signal significances are determined from $\sqrt{-2\ln(\mathcal{L}_0/\mathcal{L}_{\rm max})}$~\cite{Wilks:1938dza}, where $\mathcal{L}_{\rm max}$ and $\mathcal{L}_0$ are the likelihood values with and without a signal component, respectively. The upper limit at the 90 \% {credibility level~(C.L.)}\ on the signal yield for $\Xi_c^0 \to \Lambda\pi^0$ is obtained by solving the equation $\int_{0}^{N_{\rm UL}}{\mathcal L}(N)dN=0.9\int_{0}^{\infty}{\mathcal L}(N)dN$,
where $N$ is the assumed signal yield and $\mathcal{L}(N)$ is the maximized profiled likelihood of the fit. The upper limit, including systematic uncertainties (see Section~\ref{sec:uncer}), is calculated to be $N_{\rm UL}=454$ events {for the $\Xi_c^0 \to \Lambda \pi^0$ decay}.}

The ratios of branching fractions of $\Xi_c^0 \to \Lambda h^0$ relative to that of $\Xi_c^0 \to \Xi^- \pi^+$ are calculated using
{
\begin{multline}\label{eq:re_Br}
    \frac{{\mathcal B}(\Xi_c^0 \to \Lambda h^0)}{{\mathcal B} (\Xi_{c}^{0} \to \Xi^{-} \pi^{+})} =\\ 
    N_{\Lambda h^0} \cdot \left[\Sigma_{i}\Sigma_{j} (N_{\rm \Xi^- \pi^+}^{j} \cdot
    \frac{\epsilon_{i,\Lambda h^0}^{j}}{\epsilon_{\Xi^{-}\pi^{+}}^j} \cdot
    \frac{{\mathcal B_{i}(h^0)}}{{\mathcal B}(\Xi^{-} \to \Lambda \pi^{-})})\right]^{-1},
\end{multline}}%
where $N_{\Lambda h^0}$ and $N_{\rm \Xi^- \pi^+}$ denote the fitted {signal yields} for the signal and {normalization} modes, respectively. {The number of fitted events $N_{\Lambda h^0}$} is obtained from the fit, incorporating the contributions from different $h^0$ decay channels and datasets. The notations $\epsilon_{\Xi^{-}\pi^{+}}$ and $\epsilon_{\Lambda h^0}$ are the selection efficiencies for corresponding decay modes, ${\mathcal B} (\Xi^- \to \Lambda \pi^{-})$ is the branching fraction of $\Xi^- \to \Lambda \pi^{-}$ decay, and ${\mathcal B}(h^{0})$ denotes the product branching fractions of the respective $h^0$ decays~\cite{ParticleDataGroup:2024cfk}. {Here, $i$ indexes different $h^0$ decay channels, and $j$ indicates the quantities from Belle or Belle~II samples.} The branching fraction of $\Lambda \to p \pi^-$ {cancels} since it appears in both the numerator and denominator. These parameters are summarized in Table~\ref{tab:eff_Br}. {The 90\% C.L. upper limit on the branching ratio $\mathcal{B}(\Xi_c^0 \to \Lambda \pi^0)/\mathcal{B}(\Xi_c^0 \to \Xi^- \pi^+)$ is obtained from the mentioned upper limit equation after substituting $N$ with this branching ratio. It includes {the} systematic uncertainties detailed in Section~\ref{sec:uncer}.}

\begin{table*}[!htbp]
    \centering
    \caption{Summary of fitted signal yields ($N_{\rm fit}$), product branching fractions ($\mathcal{B}(h^0)$), and signal efficiencies in Belle ($\epsilon^{B1}$) and Belle~II ($\epsilon^{B2}$) for $\Xi_c^0$ {normalization} and signal modes. Uncertainties in $\mathcal{B}(h^0)$ are from PDG values; other uncertainties are statistical.}\label{tab:eff_Br}
    \vspace{2mm}
    \renewcommand{\arraystretch}{1.25}
    {
    \begin{tabular}{L{5.5cm} C{2cm} C{3cm} C{2.5cm} C{2.5cm}}
    \hline\hline
    \rule{0pt}{15pt}
        Decays & $N_{\rm fit}$ & $\mathcal{B}(h^0)$ (\%) & $\epsilon^{\rm B1}$ (\%) & $\epsilon^{\rm B2}$ (\%)\\[2mm] 
        \hline 
        $\Xi_c^0 \to \Xi^- \pi^+$ (Belle) & $30230\pm281$ & --- & $11.76\pm0.05$ & ---\\
        $\Xi_c^0 \to \Xi^- \pi^+$ (Belle~II) & $11579\pm161$ & --- & --- & $11.35\pm0.03$ \\\hline
        $\Xi_c^0 \to \Lambda \eta, \eta \to \gamma \gamma$ & \multirow{2}{*}{$262\pm57$} & $39.36\pm0.18$ & $3.21\pm0.02$ & $2.73\pm0.02$\\
        $\Xi_c^0 \to \Lambda \eta, \eta \to \pi^+ \pi^- \pi^0$ & & $22.75 \pm 0.25$ & $2.24\pm0.02$ & $2.72\pm0.02$\\[2mm]
        $\Xi_c^0 \to \Lambda \eta', \eta' \to \eta \pi^+ \pi^-, \eta \to \gamma \gamma$ & \multirow{3}{*}{$101\pm33$} & $16.73 \pm 0.22$ & $2.57\pm 0.02$ & $3.47 \pm 0.03$\\
        $\Xi_c^0 \to \Lambda \eta', \eta' \to \eta \pi^+ \pi^-, \eta \to \pi^+ \pi^- \pi^0$ & & $9.67\pm 0.16$ & $ 1.60\pm0.02$ & $1.78\pm0.02$ \\
        $\Xi_c^0 \to \Lambda \eta', \eta' \to \pi^+ \pi^- \gamma$ & & $29.50\pm 0.40$ & $1.64\pm0.02$ & $1.82\pm0.03$ \\[2mm]
        $\Xi_c^0 \to \Lambda \pi^0$ & $190\pm120$ & $98.83\pm0.04$ & {$2.92\pm0.03$} & {$5.15\pm0.03$}\\
        \hline\hline
    \end{tabular}
    }
\end{table*}

\section{Systematic uncertainties}\label{sec:uncer}

Systematic uncertainties in the measurements of the branching ratios include the uncertainties from the signal selection efficiencies, the MC statistics, the branching fractions of the intermediate states, and the fitting procedures. Some uncertainties arising from efficiency-related sources and branching fractions of intermediate states {cancel} when taking the ratio to the normalization mode. The total systematic uncertainties in the measurements of the branching fractions are determined by summing the individual uncertainties from {these} sources in quadrature, as detailed in Table~\ref{tab:uncer}.

The detection efficiencies determined from the simulations are corrected using multiplicative data-to-simulation ratios derived from control samples. The uncertainties associated with these correction factors are treated as systematic uncertainties. These corrections account for signal efficiency uncertainties related to tracking, PID, $\pi^0$ reconstruction, photon detection, $\pi^0$ veto, and $\Lambda$ momentum distributions. 

{The track-finding efficiency at Belle is calibrated using a single control sample, $D^{*+} \to D^0 (\to K_S^0 \pi^+ \pi^-) \pi^+$, resulting in a single, global correction and uncertainty. In contrast, efficiency correction factors and associated systematic uncertainties at Belle~II are evaluated for charged tracks by taking into account their momentum distributions from two control samples, $\bar{B}^0 \to D^{*+} (\to D^0 \pi^+) \pi^-$ and $e^+e^- \to \tau^+ \tau^-$~\cite{BelleIITrackingGroup:2020hpx}.} For a given $\Xi_c^0$ decay mode, correction factors and conservative systematic uncertainties are calculated assuming 100\% correlation among all tracks. In the case of a specific $\Xi_c^0$ signal decay mode with multiple $h^0$ decay modes, the correction factors and systematic uncertainties are weighted based on the product of signal efficiency and product branching fraction for each $h^0$ decay mode. This approach is applied for {all} uncertainty estimates, unless otherwise stated. 

{The corrections and uncertainties for} charged pion identification are obtained from the control sample of $D^{*+} \to D^0(\to K^- \pi^+) \pi^+$ at Belle~\cite{Nakano:2002jw} and combined results from {$D^{*+} \to D^0(\to K^- \pi^+) \pi^+$}, $K_S^0 \to \pi^+ \pi^-$, and $\Lambda \to p \pi^-$ at Belle~II~\cite{Belle-II:2025tpe}. {The combined control samples at Belle II provide significantly higher statistics, leading to correspondingly reduced systematic uncertainties.}

{For the Belle experiment, the $\pi^0$ reconstruction efficiency and its uncertainty are estimated using the control sample $\tau^- \to \pi^- \pi^0 \nu_\tau$, yielding Data/MC corrections of $r = (94.32 \pm 1.37)\%$, $(93.63 \pm 1.36)\%$, and $(96.41 \pm 1.40)\%$ for the $\Xi_c^0 \to \Lambda\eta~(\eta \to \pi^+ \pi^- \pi^0)$, $\Xi_c^0 \to \Lambda\eta'~(\eta' \to \pi^+ \pi^- \eta,~\eta \to \pi^+ \pi^- \pi^0)$, and $\Xi_c^0 \to \Lambda\pi^0$ channels, respectively. For Belle~II, the corresponding $\pi^0$ reconstruction efficiency and its uncertainty are measured as a function of $\pi^0$ momentum using $D^0 \to K^- \pi^+ \pi^0$, resulting in corrections of $r = (98.10 \pm 5.30)\%$, $(95.01 \pm 4.72)\%$, and $(103.32 \pm 4.27)\%$ for the same three channels. These values are subsequently weighted by the branching fractions and detection efficiencies of the corresponding signal channels to obtain the final corrections and uncertainties.}

Photon reconstruction uncertainties are obtained from radiative Bhabha and muon-pair control samples at Belle and Belle~II, respectively. {Belle adopts a conservative, fixed uncertainty of 2\% per photon. In contrast, Belle II provides momentum- and angle-dependent corrections and uncertainties for the photons.} Notably, uncertainties in photon reconstruction are already accounted for within the $\pi^0$ reconstruction uncertainties. 

{The uncertainty associated with the $\pi^0$ veto, which relies on two photons, is expected to be smaller for Belle~II due to its more accurate simulation in MC samples. This is confirmed using the control channel $\eta \to \gamma\gamma$, from which we obtain $\pi^0$ veto uncertainties to be $2.23\%$ ($1.02\%$) for $\Xi_c^0 \to \Lambda\eta$ decay and $0.71\%$ ($0.34\%$) for $\Xi_c^0 \to \Lambda\eta'$ decay in Belle (Belle~II), consistent with the better consistency between the MC simulation and data at Belle~II.}

To quantify the impact arising from the differences in the $\Lambda$ momentum distributions between the {normalization} mode and each signal mode, we reweight the $\Lambda$ momentum distribution of {the normalization} mode so that it matches that of the corresponding signal mode. {The relative deviations of reweighted results from nominal branching ratio are assigned as the systematic uncertainties associated with the difference of $\Lambda$ momentum distribution between the signal channels and normalization mode.} 

Signal MC samples are weighted according to the efficiency-corrected $x_p$ distribution of the normalization mode from data to ensure good agreement between data and MC simulations. The efficiency-corrected $x_p$ distribution is obtained by fitting the $M(\Xi^- \pi^+)$ distribution in each $x_p$ bin of data, accounting for efficiency in each bin. {The uncertainty associated with $x_p$ correction is negligible in this analysis.} 

The systematic uncertainty arising from {the} limited {MC statistics in efficiency calculation} is evaluated using a binomial estimation method.

The systematic uncertainties due to intermediate branching fractions are assigned based on the uncertainties of the world-average values. {For a specific $\Xi_c^0$ signal decay mode reconstructed via multiple $h^0$ decay modes and using combined Belle and Belle~II samples, the branching fraction of each $h^0$ decay mode is varied independently by $\pm$1$\sigma$. The resulting deviation from the nominal value is adopted as the corresponding systematic uncertainty. Finally, the uncertainties are combined in quadrature among the different $h^0$ decay modes and assigned as the uncertainties of this $\Xi_c^0$ decay mode.}

The 18.89\% uncertainty on $\mathcal{B}(\Xi_c^0 \to \Xi^- \pi^+)$~\cite{ParticleDataGroup:2024cfk} is treated as an independent systematic uncertainty in the {measurements} of the absolute branching fractions. 

The systematic uncertainty associated with the fit procedure is estimated by comparing the $\Xi_c^0$ signal yield in the nominal fit and in modified fits which {include} (i) changing the fit range by $\pm10\%$, (ii) changing to {the third-order} Chebyshev function for background fit, {(iii) changing the expected ratio of the signal yields for each $\Lambda h^0$ mode in the simultaneous fits by considering the luminosity uncertainties of Belle and Belle~II data}, {and (iv) allowing the signal width to be free in the fit for the $\Lambda \eta$ and $\Lambda \eta'$ modes}. {For each of the four fit modifications, an ensemble of pseudo-experiments is generated and fitted. The resulting signal yields of each ensemble are Gaussian-distributed; the offset of each Gaussian mean from the nominal-fit yield is taken as the systematic uncertainty associated with that modification.} The fit uncertainties for $\Xi_c^0 \to \Lambda \eta/\Lambda \eta'$ are calculated by adding the uncertainties from {four} modifications in quadrature. For the unobserved decay $\Xi_c^0 \to \Lambda \pi^0$, we calculate the upper limit for {all the possible combinations and} take the highest value as the 90\% C.L.\ upper limit. Then, the likelihood with that most conservative upper limit is convolved with a Gaussian function whose width is equal to the corresponding total multiplicative uncertainty summarized in Table~\ref{tab:uncer}. We estimate the fit uncertainties for the signal and normalization modes separately, and the uncertainties for the normalization mode are determined to be 0.17\% and {0.33\%} for Belle and Belle~II samples, respectively. Finally, the fit uncertainties of the signal and normalization modes are added in quadrature to obtain the total fit uncertainty.

\begin{table*}[!htbp]
    \centering
    \renewcommand{\arraystretch}{1.2}
    \caption{Fractional systematic uncertainties (\%) on the {branching ratios} from different sources. {Systematic uncertainties associated with the fitting procedures are treated as multiplicative for $\Xi_c^0 \to \Lambda \eta/\Lambda \eta'$ and as additive for the unobserved mode $\Xi_c^0 \to \Lambda \pi^0$.} The total uncertainties are calculated by first summing the uncertainties from different sources in quadrature for Belle and Belle~II separately, and then deriving the results from the luminosity-weighted average of these sums.}\label{tab:uncer}
    \begin{tabular}{L{4cm} C{1.5cm} C{1.5cm} C{1.5cm} C{1.5cm} C{1.5cm} C{1.5cm}}
        \hline\hline
        \rule{0pt}{15pt}
        \multirow{2}{*}{Source} & \multicolumn{2}{c}{\large $\frac{{\mathcal{B}(\Xi_c^0 \to \Lambda \eta)}}{{\mathcal{B}(\Xi_c^0 \to \Xi^- \pi^+)}}$} & \multicolumn{2}{c}{\large $\frac{{\mathcal{B}(\Xi_c^0 \to \Lambda \eta')}}{{\mathcal{B}(\Xi_c^0 \to \Xi^- \pi^+)}}$} & \multicolumn{2}{c}{\large $\frac{{\mathcal{B}(\Xi_c^0 \to \Lambda \pi^0)}}{{\mathcal{B}(\Xi_c^0 \to \Xi^- \pi^+)}}$}\\
        & Belle & Belle~II & Belle & Belle~II & Belle & Belle~II\\
        \hline
        Tracking efficiency & 0.77\% & 1.07\% & 1.51\% & 2.13\% & 0.64\% & 0.80\%\\
        $\pi^+$ PID & 1.49\% & 0.21\% & 2.24\% & 0.24\% & 1.39\% & 0.20\%\\
        $\pi^0$ reconstruction & 0.24\% & 1.67\% & 0.18\% & 0.65\% & 1.45\% & 4.13\%\\
        Photon reconstruction & 3.35\% & 0.84\% & 2.54\% & 0.98\% & --- & ---\\
        $\pi^0$ veto & 2.23\% & 1.02\% & 0.71\% & 0.34\% & --- & ---\\
        $\Lambda$ momentum & {0.56\%} & {0.34\%} & {0.55\%} & {0.68\%} & {0.20\%} & {0.82\%}\\ 
        MC sample size & 1.08\% & 0.82\% & 1.15\% & 1.08\% & 0.67\% & 0.45\%\\
        Intermediate states ${\mathcal B}$ & 0.47\% & 0.47\% & 0.85\% & 0.85\% & 0.06\% & 0.06\%\\
        Fit {procedure} & 5.54\% & 5.54\% & {4.83\%} & {4.84\%} & {0.17\%} & {0.33\%}\\
        \hline
        Total & \multicolumn{2}{c}{{5.35\%}} & \multicolumn{2}{c}{{4.77\%}} & \multicolumn{2}{c}{{2.04\%}}\\
        \hline
        {Normalization} mode ${\mathcal B}$ & \multicolumn{6}{c}{18.89\%}\\
        \hline\hline
    \end{tabular}
\end{table*}

\section{Result and discussion}
We present the first measurements of the branching fractions of the singly Cabibbo-suppressed decays $\Xi_c^0 \to \Lambda \eta$, $\Xi_c^0 \to \Lambda \eta'$, and $\Xi_c^0 \to \Lambda \pi^0$, using the combined data samples from Belle and Belle~II. {We report the first observation of the decay $\Xi_c^0 \to \Lambda \eta$, and the first evidence of the decay $\Xi_c^0 \to \Lambda \eta'$. The measured branching ratios of these two decays are
\begin{equation}
    \begin{aligned}
        \frac{\mathcal{B}(\Xi_c^0 \to \Lambda \eta)}{\mathcal{B}(\Xi_c^0 \to \Xi^- \pi^+)}&= (4.16 \pm 0.91 \pm {0.23})\%\nonumber
    \end{aligned}
\end{equation}}
and
{
\begin{equation}
    \begin{aligned}
        \frac{\mathcal{B}(\Xi_c^0 \to \Lambda \eta')}{\mathcal{B}(\Xi_c^0 \to \Xi^- \pi^+)}= (2.48 \pm 0.82 \pm {0.12})\%. \nonumber
    \end{aligned}
\end{equation}}%
{We find no evidence of the decay $\Xi_c^0 \to \Lambda \pi^0$ and set an upper limit at the 90\% C.L.\ of}
{
\begin{equation}
    \begin{aligned}
        \frac{\mathcal{B}(\Xi_c^0 \to \Lambda \pi^0)}{\mathcal{B}(\Xi_c^0 \to \Xi^- \pi^+)}&< {3.5\%}.\nonumber 
    \end{aligned}
\end{equation}}%
Branching ratios obtained from independent fits to Belle and Belle~II data {are consistent with those obtained from simultaneous fits.} 

Taking $\mathcal{B}(\Xi_c^0 \to \Xi^- \pi^+)=(1.43\pm0.27)$\%~\cite{ParticleDataGroup:2024cfk} {with its uncertainty included in the total multiplicative systematic uncertainty}, the measured absolute branching fractions of the decays $\Xi_c^0 \to \Lambda \eta$ and $\Xi_c^0 \to \Lambda \eta'$ are
{
\begin{equation}
    \begin{aligned}
        \mathcal{B}(\Xi_c^0 \to \Lambda \eta)&= (5.95 \pm 1.30 \pm {0.32} \pm 1.13) \times 10^{-4}\nonumber
    \end{aligned}
\end{equation}}
and
{
\begin{equation}
    \begin{aligned}
        \mathcal{B}(\Xi_c^0 \to \Lambda \eta')&= (3.55 \pm 1.17 \pm {0.17} \pm 0.68) \times 10^{-4}. \nonumber
    \end{aligned}
\end{equation}}%
Here, the uncertainties are statistical, systematic, and from $\mathcal{B}(\Xi_c^0 \to \Xi^- \pi^+)$, respectively. We also obtain the 90\% C.L.\ upper limit on the absolute branching fraction of the decay $\Xi_c^0 \to \Lambda \pi^0$:
{
\begin{equation}
    \begin{aligned}
        \mathcal{B}(\Xi_c^0 \to \Lambda \pi^0)&< {5.2} \times 10^{-4}.\nonumber\\
    \end{aligned}
\end{equation}}%

\begin{figure*}[!htbp]
    \begin{minipage}[c]{0.65\linewidth}
        \raggedright
        \includegraphics[width=\textwidth]{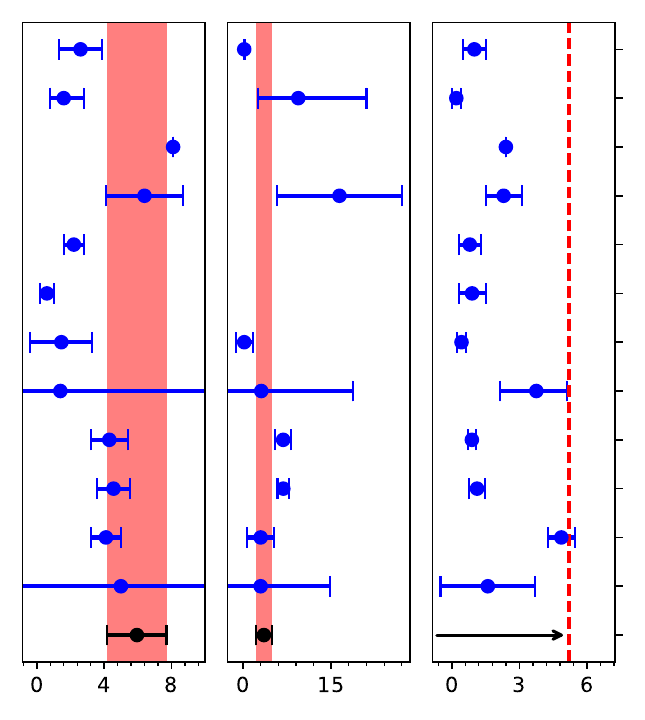}\\[-\baselineskip]
        \text{\small \hspace{0.5cm} $\mathcal{B}(\Xi_c^0 \to \Lambda \eta) (\times 10^{-4}$) \hspace{0.5cm} $\mathcal{B}(\Xi_c^0 \to \Lambda \eta') (\times 10^{-4}$) \hspace{0.5cm} $\mathcal{B}(\Xi_c^0 \to \Lambda \pi^0) (\times 10^{-4}$)}\\   
    \end{minipage}
    \begin{minipage}[c]{0.25\linewidth}
        \large
        \raggedright
        Zhao ${\it et~al.}$~\cite{Zhao:2018mov} \\[4.0mm]
        Geng ${\it et~al.}$~\cite{Geng:2018plk}\\[4.0mm]
        Zou ${\it et~al.}$~\cite{Zou:2019kzq}\\[4.0mm]
        Geng ${\it et~al.}$\cite{Geng:2019xbo}\\[4.0mm]         
        Hsiao ${\it et~al.}$ (I)~\cite{Hsiao:2021nsc}\\[4.0mm]
        Hsiao ${\it et~al.}$ (II)~\cite{Hsiao:2021nsc}\\[4.0mm]
        Zhong ${\it et~al.}$ (I)~\cite{Zhong:2022exp}\\[4.0mm]
        Zhong ${\it et~al.}$ (II)~\cite{Zhong:2022exp}\\[4.0mm]
        Geng ${\it et~al.}$~\cite{Geng:2023pkr}\\[4.0mm]
        Zhong ${\it et~al.}$~\cite{Zhong:2024zme}\\[4.0mm]
        Xing ${\it et~al.}$ (I)~\cite{Xing:2024nvg}\\[4.0mm]
        Xing ${\it et~al.}$ (II)~\cite{Xing:2024nvg}\\[4.0mm]
        Belle and Belle~II combined measurements \\
        \vspace{3mm}
    \end{minipage}
    \caption{Comparison of the branching fractions $\mathcal{B}(\Xi_c^0 \to \Lambda \eta)$, $\mathcal{B}(\Xi_c^0 \to \Lambda \eta')$, and of the measured 90\% C.L.\ upper limit on $\mathcal{B}(\Xi_c^0 \to \Lambda \pi^0)$ with their respective theoretical predictions~\cite{Zhao:2018mov,Geng:2018plk,Zou:2019kzq,Geng:2019xbo,Hsiao:2021nsc,Zhong:2022exp,Geng:2023pkr,Zhong:2024zme,Xing:2024nvg}. Dots with error bars represent central values and their uncertainties; those without indicate a lack of theoretical uncertainty. Missing dots signify the absence of theoretical predictions for that decay mode. Some predictions exhibit large errors that are not fully captured within the present scale. The black dots with error bars denote the measured absolute branching fractions of decays $\Xi_c^0 \to \Lambda \eta$ and $\Xi_c^0 \to \Lambda \eta'$. {The red vertical bands indicate the $\pm1\sigma$ intervals which are dominated by the systematic uncertainty of $\mathcal{B}(\Xi_c^0 \to \Xi^- \pi^+)$.} The red dashed line and the black arrow indicate the measured 90\% C.L. upper limits of the $\mathcal{B}(\Xi_c^0 \to \Lambda \pi^0)$. For refs.~\cite{Hsiao:2021nsc,Zhong:2022exp}, (I) indicates the predicted value based on the SU(3)$_{F}$ symmetry, while (II) takes into account the breaking SU(3)$_{F}$ symmetry. For the Ref.~\cite{Xing:2024nvg}, (I) and (II) denote the predicted values based on real form factors alone and those incorporating complex form factors, respectively.}
    \label{fig:data_br}
\end{figure*}

Figure~\ref{fig:data_br} compares measured $\Xi_c^0 \to \Lambda \eta$ and $\Xi_c^0 \to \Lambda \eta'$ branching fractions and 90\% C.L. upper limit of $\mathcal{B}(\Xi_c^0 \to \Lambda \pi^0)$ with twelve theoretical predictions~\cite{Zhao:2018mov,Geng:2018plk,Zou:2019kzq,Geng:2019xbo,Hsiao:2021nsc,Zhong:2022exp,Geng:2023pkr,Zhong:2024zme,Xing:2024nvg}. Three SU(3)$_F$-based predictions~\cite{Zhong:2022exp,Xing:2024nvg} agree with our measured $\mathcal{B}(\Xi_c^0 \to \Lambda \eta)$ and $\mathcal{B}(\Xi_c^0 \to \Lambda \eta')$ within 1$\sigma$. All twelve predictions {are} within 3$\sigma$ of {the} measured branching fractions for both $\Xi_c^0 \to \Lambda \eta$ and $\Xi_c^0 \to \Lambda \eta'$, and lie below the 90\% C.L.\ upper limit on $\mathcal{B}(\Xi_c^0 \to \Lambda \pi^0)$. The ratios $\mathcal{B}(\Xi_c^0 \to \Lambda \eta)/\mathcal{B}(\Xi_c^0 \to \Xi^- \pi^+)$, $\mathcal{B}(\Xi_c^0 \to \Lambda \eta')/\mathcal{B}(\Xi_c^0 \to \Xi^- \pi^+)$, and $\mathcal{B}(\Xi_c^0 \to \Lambda \pi^0)/\mathcal{B}(\Xi_c^0 \to \Xi^- \pi^+)$ are independent of the $\Xi_c^0$ absolute branching fraction scale and may also be compared to theoretical models.

This work, based on data collected using the Belle II detector, which was built and commissioned prior to March 2019,
and data collected using the Belle detector, which was operated until June 2010,
was supported by
Higher Education and Science Committee of the Republic of Armenia Grant No.~23LCG-1C011;
Australian Research Council and Research Grants
No.~DP200101792, 
No.~DP210101900, 
No.~DP210102831, 
No.~DE220100462, 
No.~LE210100098, 
and
No.~LE230100085; 
Austrian Federal Ministry of Education, Science and Research,
Austrian Science Fund (FWF) Grants
DOI:~10.55776/P34529,
DOI:~10.55776/J4731,
DOI:~10.55776/J4625,
DOI:~10.55776/M3153,
and
DOI:~10.55776/PAT1836324,
and
Horizon 2020 ERC Starting Grant No.~947006 ``InterLeptons'';
Natural Sciences and Engineering Research Council of Canada, Digital Research Alliance of Canada, and Canada Foundation for Innovation;
National Key R\&D Program of China under Contract No.~2024YFA1610503,
and
No.~2024YFA1610504
National Natural Science Foundation of China and Research Grants
No.~11575017,
No.~11761141009,
No.~11705209,
No.~11975076,
No.~12135005,
No.~12150004,
No.~12161141008,
No.~12405099,
No.~12475093,
and
No.~12175041,
and Shandong Provincial Natural Science Foundation Project~ZR2022JQ02;
the Czech Science Foundation Grant No. 22-18469S,  Regional funds of EU/MEYS: OPJAK
FORTE CZ.02.01.01/00/22\_008/0004632 
and
Charles University Grant Agency project No. 246122;
European Research Council, Seventh Framework PIEF-GA-2013-622527,
Horizon 2020 ERC-Advanced Grants No.~267104 and No.~884719,
Horizon 2020 ERC-Consolidator Grant No.~819127,
Horizon 2020 Marie Sklodowska-Curie Grant Agreement No.~700525 ``NIOBE''
and
No.~101026516,
and
Horizon 2020 Marie Sklodowska-Curie RISE project JENNIFER2 Grant Agreement No.~822070 (European grants);
L'Institut National de Physique Nucl\'{e}aire et de Physique des Particules (IN2P3) du CNRS
and
L'Agence Nationale de la Recherche (ANR) under Grant No.~ANR-23-CE31-0018 (France);
BMFTR, DFG, HGF, MPG, and AvH Foundation (Germany);
Department of Atomic Energy under Project Identification No.~RTI 4002,
Department of Science and Technology,
and
UPES SEED funding programs
No.~UPES/R\&D-SEED-INFRA/17052023/01 and
No.~UPES/R\&D-SOE/20062022/06 (India);
Israel Science Foundation Grant No.~2476/17,
U.S.-Israel Binational Science Foundation Grant No.~2016113, and
Israel Ministry of Science Grant No.~3-16543;
Istituto Nazionale di Fisica Nucleare and the Research Grants BELLE2,
and
the ICSC – Centro Nazionale di Ricerca in High Performance Computing, Big Data and Quantum Computing, funded by European Union – NextGenerationEU;
Japan Society for the Promotion of Science, Grant-in-Aid for Scientific Research Grants
No.~16H03968,
No.~16H03993,
No.~16H06492,
No.~16K05323,
No.~17H01133,
No.~17H05405,
No.~18K03621,
No.~18H03710,
No.~18H05226,
No.~19H00682, 
No.~20H05850,
No.~20H05858,
No.~22H00144,
No.~22K14056,
No.~22K21347,
No.~23H05433,
No.~26220706,
and
No.~26400255,
and
the Ministry of Education, Culture, Sports, Science, and Technology (MEXT) of Japan;  
National Research Foundation (NRF) of Korea Grants
No.~2021R1-F1A-1064008, 
No.~2022R1-A2C-1003993,
No.~2022R1-A2C-1092335,
No.~RS-2016-NR017151,
No.~RS-2018-NR031074,
No.~RS-2021-NR060129,
No.~RS-2023-00208693,
No.~RS-2024-00354342
and
No.~RS-2025-02219521,
Radiation Science Research Institute,
Foreign Large-Size Research Facility Application Supporting project,
the Global Science Experimental Data Hub Center, the Korea Institute of Science and
Technology Information (K25L2M2C3 ) 
and
KREONET/GLORIAD;
Universiti Malaya RU grant, Akademi Sains Malaysia, and Ministry of Education Malaysia;
Frontiers of Science Program Contracts
No.~FOINS-296,
No.~CB-221329,
No.~CB-236394,
No.~CB-254409,
and
No.~CB-180023, and SEP-CINVESTAV Research Grant No.~237 (Mexico);
the Polish Ministry of Science and Higher Education and the National Science Center;
the Ministry of Science and Higher Education of the Russian Federation
and
the HSE University Basic Research Program, Moscow;
University of Tabuk Research Grants
No.~S-0256-1438 and No.~S-0280-1439 (Saudi Arabia), and
Researchers Supporting Project number (RSPD2025R873), King Saud University, Riyadh,
Saudi Arabia;
Slovenian Research Agency and Research Grants
No.~J1-50010
and
No.~P1-0135;
Ikerbasque, Basque Foundation for Science,
State Agency for Research of the Spanish Ministry of Science and Innovation through Grant No. PID2022-136510NB-C33, Spain,
Agencia Estatal de Investigacion, Spain
Grant No.~RYC2020-029875-I
and
Generalitat Valenciana, Spain
Grant No.~CIDEGENT/2018/020;
the Swiss National Science Foundation;
The Knut and Alice Wallenberg Foundation (Sweden), Contracts No.~2021.0174, No.~2021.0299, and No.~2023.0315;
National Science and Technology Council,
and
Ministry of Education (Taiwan);
Thailand Center of Excellence in Physics;
TUBITAK ULAKBIM (Turkey);
National Research Foundation of Ukraine, Project No.~2020.02/0257,
and
Ministry of Education and Science of Ukraine;
the U.S. National Science Foundation and Research Grants
No.~PHY-1913789 
and
No.~PHY-2111604, 
and the U.S. Department of Energy and Research Awards
No.~DE-AC06-76RLO1830, 
No.~DE-SC0007983, 
No.~DE-SC0009824, 
No.~DE-SC0009973, 
No.~DE-SC0010007, 
No.~DE-SC0010073, 
No.~DE-SC0010118, 
No.~DE-SC0010504, 
No.~DE-SC0011784, 
No.~DE-SC0012704, 
No.~DE-SC0019230, 
No.~DE-SC0021274, 
No.~DE-SC0021616, 
No.~DE-SC0022350, 
No.~DE-SC0023470; 
and
the Vietnam Academy of Science and Technology (VAST) under Grants
No.~NVCC.05.02/25-25
and
No.~DL0000.05/26-27.

These acknowledgements are not to be interpreted as an endorsement of any statement made
by any of our institutes, funding agencies, governments, or their representatives.

We thank the SuperKEKB team for delivering high-luminosity collisions;
the KEK cryogenics group for the efficient operation of the detector solenoid magnet and IBBelle on site;
the KEK Computer Research Center for on-site computing support; the NII for SINET6 network support;
and the raw-data centers hosted by BNL, DESY, GridKa, IN2P3, INFN, 
PNNL/EMSL, 
and the University of Victoria.

Numerical data corresponding to the results presented are available as HEPData.
The full Belle II data are not publicly available.
The collaboration will consider requests for access to the data that support this article.


\begin{thebibliography}{54}%
\makeatletter
\providecommand \@ifxundefined [1]{%
 \@ifx{#1\undefined}
}%
\providecommand \@ifnum [1]{%
 \ifnum #1\expandafter \@firstoftwo
 \else \expandafter \@secondoftwo
 \fi
}%
\providecommand \@ifx [1]{%
 \ifx #1\expandafter \@firstoftwo
 \else \expandafter \@secondoftwo
 \fi
}%
\providecommand \natexlab [1]{#1}%
\providecommand \enquote  [1]{``#1''}%
\providecommand \bibnamefont  [1]{#1}%
\providecommand \bibfnamefont [1]{#1}%
\providecommand \citenamefont [1]{#1}%
\providecommand \href@noop [0]{\@secondoftwo}%
\providecommand \href [0]{\begingroup \@sanitize@url \@href}%
\providecommand \@href[1]{\@@startlink{#1}\@@href}%
\providecommand \@@href[1]{\endgroup#1\@@endlink}%
\providecommand \@sanitize@url [0]{\catcode `\\12\catcode `\$12\catcode
  `\&12\catcode `\#12\catcode `\^12\catcode `\_12\catcode `\%12\relax}%
\providecommand \@@startlink[1]{}%
\providecommand \@@endlink[0]{}%
\providecommand \url  [0]{\begingroup\@sanitize@url \@url }%
\providecommand \@url [1]{\endgroup\@href {#1}{\urlprefix }}%
\providecommand \urlprefix  [0]{URL }%
\providecommand \Eprint [0]{\href }%
\providecommand \doibase [0]{https://doi.org/}%
\providecommand \selectlanguage [0]{\@gobble}%
\providecommand \bibinfo  [0]{\@secondoftwo}%
\providecommand \bibfield  [0]{\@secondoftwo}%
\providecommand \translation [1]{[#1]}%
\providecommand \BibitemOpen [0]{}%
\providecommand \bibitemStop [0]{}%
\providecommand \bibitemNoStop [0]{.\EOS\space}%
\providecommand \EOS [0]{\spacefactor3000\relax}%
\providecommand \BibitemShut  [1]{\csname bibitem#1\endcsname}%
\let\auto@bib@innerbib\@empty
\bibitem [{\citenamefont {Chau}\ \emph {et~al.}(1996)\citenamefont {Chau},
  \citenamefont {Cheng},\ and\ \citenamefont {Tseng}}]{Chau:1995gk}%
  \BibitemOpen
  \bibfield  {author} {\bibinfo {author} {\bibfnamefont {L.~L.}\ \bibnamefont
  {Chau}}, \bibinfo {author} {\bibfnamefont {H.~Y.}\ \bibnamefont {Cheng}},\
  and\ \bibinfo {author} {\bibfnamefont {B.}~\bibnamefont {Tseng}},\ }\href
  {https://doi.org/10.1103/PhysRevD.54.2132} {\bibfield  {journal} {\bibinfo
  {journal} {{Phys. Rev. D}}\ }\textbf {\bibinfo {volume} {54}},\ \bibinfo
  {pages} {2132} (\bibinfo {year} {1996})}\BibitemShut {NoStop}%
\bibitem [{\citenamefont {Cheng}(2022{\natexlab{a}})}]{Cheng:2021qpd}%
  \BibitemOpen
  \bibfield  {author} {\bibinfo {author} {\bibfnamefont {H.~Y.}\ \bibnamefont
  {Cheng}},\ }\href {https://doi.org/10.1016/j.cjph.2022.06.021} {\bibfield
  {journal} {\bibinfo  {journal} {Chin. J. Phys.}\ }\textbf {\bibinfo {volume}
  {78}},\ \bibinfo {pages} {324} (\bibinfo {year}
  {2022}{\natexlab{a}})}\BibitemShut {NoStop}%
\bibitem [{\citenamefont {Cheng}\ \emph {et~al.}(2018)\citenamefont {Cheng},
  \citenamefont {Kang},\ and\ \citenamefont {Xu}}]{Cheng:2018hwl}%
  \BibitemOpen
  \bibfield  {author} {\bibinfo {author} {\bibfnamefont {H.~Y.}\ \bibnamefont
  {Cheng}}, \bibinfo {author} {\bibfnamefont {X.~W.}\ \bibnamefont {Kang}},\
  and\ \bibinfo {author} {\bibfnamefont {F.}~\bibnamefont {Xu}},\ }\href
  {https://doi.org/10.1103/PhysRevD.97.074028} {\bibfield  {journal} {\bibinfo
  {journal} {{Phys. Rev. D}}\ }\textbf {\bibinfo {volume} {97}},\ \bibinfo
  {pages} {074028} (\bibinfo {year} {2018})}\BibitemShut {NoStop}%
\bibitem [{\citenamefont {Navas}\ \emph {et~al.}(2024)\citenamefont {Navas}
  \emph {et~al.}}]{ParticleDataGroup:2024cfk}%
  \BibitemOpen
  \bibfield  {author} {\bibinfo {author} {\bibfnamefont {S.}~\bibnamefont
  {Navas}} \emph {et~al.} (\bibinfo {collaboration} {Particle Data Group}),\
  }\href {https://doi.org/10.1103/PhysRevD.110.030001} {\bibfield  {journal}
  {\bibinfo  {journal} {{Phys. Rev. D}}\ }\textbf {\bibinfo {volume} {110}},\
  \bibinfo {pages} {030001} (\bibinfo {year} {2024})}\BibitemShut {NoStop}%
\bibitem [{\citenamefont {Zupanc}\ \emph {et~al.}(2014)\citenamefont {Zupanc}
  \emph {et~al.}}]{Belle:2013jfq}%
  \BibitemOpen
  \bibfield  {author} {\bibinfo {author} {\bibfnamefont {A.}~\bibnamefont
  {Zupanc}} \emph {et~al.} (\bibinfo {collaboration} {Belle collaboration}),\
  }\href {https://doi.org/10.1103/PhysRevLett.113.042002} {\bibfield  {journal}
  {\bibinfo  {journal} {{Phys. Rev. Lett.}}\ }\textbf {\bibinfo {volume}
  {113}},\ \bibinfo {pages} {042002} (\bibinfo {year} {2014})}\BibitemShut
  {NoStop}%
\bibitem [{\citenamefont {Ablikim}\ \emph {et~al.}(2016)\citenamefont {Ablikim}
  \emph {et~al.}}]{BESIII:2015bjk}%
  \BibitemOpen
  \bibfield  {author} {\bibinfo {author} {\bibfnamefont {M.}~\bibnamefont
  {Ablikim}} \emph {et~al.} (\bibinfo {collaboration} {BESIII collaboration}),\
  }\href {https://doi.org/10.1103/PhysRevLett.116.052001} {\bibfield  {journal}
  {\bibinfo  {journal} {{Phys. Rev. Lett.}}\ }\textbf {\bibinfo {volume}
  {116}},\ \bibinfo {pages} {052001} (\bibinfo {year} {2016})}\BibitemShut
  {NoStop}%
\bibitem [{\citenamefont {Li}\ \emph {et~al.}(2019{\natexlab{a}})\citenamefont
  {Li} \emph {et~al.}}]{Belle:2019bgi}%
  \BibitemOpen
  \bibfield  {author} {\bibinfo {author} {\bibfnamefont {Y.~B.}\ \bibnamefont
  {Li}} \emph {et~al.} (\bibinfo {collaboration} {Belle collaboration}),\
  }\href {https://doi.org/10.1103/PhysRevD.100.031101} {\bibfield  {journal}
  {\bibinfo  {journal} {{Phys. Rev. D}}\ }\textbf {\bibinfo {volume} {100}},\
  \bibinfo {pages} {031101} (\bibinfo {year} {2019}{\natexlab{a}})}\BibitemShut
  {NoStop}%
\bibitem [{\citenamefont {Li}\ \emph {et~al.}(2019{\natexlab{b}})\citenamefont
  {Li} \emph {et~al.}}]{Belle:2018kzz}%
  \BibitemOpen
  \bibfield  {author} {\bibinfo {author} {\bibfnamefont {Y.~B.}\ \bibnamefont
  {Li}} \emph {et~al.} (\bibinfo {collaboration} {Belle collaboration}),\
  }\href {https://doi.org/10.1103/PhysRevLett.122.082001} {\bibfield  {journal}
  {\bibinfo  {journal} {{Phys. Rev. Lett.}}\ }\textbf {\bibinfo {volume}
  {122}},\ \bibinfo {pages} {082001} (\bibinfo {year}
  {2019}{\natexlab{b}})}\BibitemShut {NoStop}%
\bibitem [{\citenamefont {Aaij}\ \emph {et~al.}(2018)\citenamefont {Aaij} \emph
  {et~al.}}]{LHCb:2017xtf}%
  \BibitemOpen
  \bibfield  {author} {\bibinfo {author} {\bibfnamefont {R.}~\bibnamefont
  {Aaij}} \emph {et~al.} (\bibinfo {collaboration} {LHCb collaboration}),\
  }\href {https://doi.org/10.1007/JHEP03(2018)043} {\bibfield  {journal}
  {\bibinfo  {journal} {{J. High Energy Phys.}}\ }\textbf {\bibinfo {volume}
  {03}},\ \bibinfo {pages} {043} (\bibinfo {year} {2018})}\BibitemShut
  {NoStop}%
\bibitem [{\citenamefont {Li}\ \emph {et~al.}(2021)\citenamefont {Li} \emph
  {et~al.}}]{Belle:2021mvw}%
  \BibitemOpen
  \bibfield  {author} {\bibinfo {author} {\bibfnamefont {S.~X.}\ \bibnamefont
  {Li}} \emph {et~al.} (\bibinfo {collaboration} {Belle collaboration}),\
  }\href {https://doi.org/10.1103/PhysRevD.103.072004} {\bibfield  {journal}
  {\bibinfo  {journal} {Phys. Rev. D}\ }\textbf {\bibinfo {volume} {103}},\
  \bibinfo {pages} {072004} (\bibinfo {year} {2021})}\BibitemShut {NoStop}%
\bibitem [{\citenamefont {Aaij}\ \emph {et~al.}(2019)\citenamefont {Aaij} \emph
  {et~al.}}]{LHCb:2019nxp}%
  \BibitemOpen
  \bibfield  {author} {\bibinfo {author} {\bibfnamefont {R.}~\bibnamefont
  {Aaij}} \emph {et~al.} (\bibinfo {collaboration} {LHCb collaboration}),\
  }\href {https://doi.org/10.1007/JHEP04(2019)084} {\bibfield  {journal}
  {\bibinfo  {journal} {{J. High Energy Phys.}}\ }\textbf {\bibinfo {volume}
  {04}},\ \bibinfo {pages} {084} (\bibinfo {year} {2019})}\BibitemShut
  {NoStop}%
\bibitem [{\citenamefont {Adachi}\ \emph
  {et~al.}(2025{\natexlab{a}})\citenamefont {Adachi} \emph
  {et~al.}}]{Belle:2024xcs}%
  \BibitemOpen
  \bibfield  {author} {\bibinfo {author} {\bibfnamefont {I.}~\bibnamefont
  {Adachi}} \emph {et~al.} (\bibinfo {collaboration} {Belle and Belle II
  collaboration}),\ }\href {https://doi.org/10.1007/JHEP03(2025)061} {\bibfield
   {journal} {\bibinfo  {journal} {{J. High Energy Phys.}}\ }\textbf {\bibinfo
  {volume} {03}},\ \bibinfo {pages} {061} (\bibinfo {year}
  {2025}{\natexlab{a}})}\BibitemShut {NoStop}%
\bibitem [{\citenamefont {Li}\ \emph {et~al.}(2022)\citenamefont {Li} \emph
  {et~al.}}]{Belle:2021avh}%
  \BibitemOpen
  \bibfield  {author} {\bibinfo {author} {\bibfnamefont {Y.}~\bibnamefont {Li}}
  \emph {et~al.} (\bibinfo {collaboration} {Belle collaboration}),\ }\href
  {https://doi.org/10.1103/PhysRevD.105.L011102} {\bibfield  {journal}
  {\bibinfo  {journal} {{Phys. Rev. D}}\ }\textbf {\bibinfo {volume} {105}},\
  \bibinfo {pages} {L011102} (\bibinfo {year} {2022})}\BibitemShut {NoStop}%
\bibitem [{\citenamefont {Adachi}\ \emph {et~al.}(2024)\citenamefont {Adachi}
  \emph {et~al.}}]{Belle:2024ikp}%
  \BibitemOpen
  \bibfield  {author} {\bibinfo {author} {\bibfnamefont {I.}~\bibnamefont
  {Adachi}} \emph {et~al.} (\bibinfo {collaboration} {Belle and Belle~II
  collaboration}),\ }\href {https://doi.org/10.1007/JHEP10(2024)045} {\bibfield
   {journal} {\bibinfo  {journal} {{J. High Energy Phys.}}\ }\textbf {\bibinfo
  {volume} {10}},\ \bibinfo {pages} {045} (\bibinfo {year} {2024})}\BibitemShut
  {NoStop}%
\bibitem [{\citenamefont {Zhao}\ \emph {et~al.}(2020)\citenamefont {Zhao},
  \citenamefont {Wang}, \citenamefont {Hsiao},\ and\ \citenamefont
  {Yu}}]{Zhao:2018mov}%
  \BibitemOpen
  \bibfield  {author} {\bibinfo {author} {\bibfnamefont {H.~J.}\ \bibnamefont
  {Zhao}}, \bibinfo {author} {\bibfnamefont {Y.~L.}\ \bibnamefont {Wang}},
  \bibinfo {author} {\bibfnamefont {Y.~K.}\ \bibnamefont {Hsiao}},\ and\
  \bibinfo {author} {\bibfnamefont {Y.}~\bibnamefont {Yu}},\ }\href
  {https://doi.org/10.1007/JHEP02(2020)165} {\bibfield  {journal} {\bibinfo
  {journal} {{J. High Energy Phys.}}\ }\textbf {\bibinfo {volume} {02}},\
  \bibinfo {pages} {165} (\bibinfo {year} {2020})}\BibitemShut {NoStop}%
\bibitem [{\citenamefont {Hsiao}\ \emph {et~al.}(2022)\citenamefont {Hsiao},
  \citenamefont {Wang},\ and\ \citenamefont {Zhao}}]{Hsiao:2021nsc}%
  \BibitemOpen
  \bibfield  {author} {\bibinfo {author} {\bibfnamefont {Y.~K.}\ \bibnamefont
  {Hsiao}}, \bibinfo {author} {\bibfnamefont {Y.~L.}\ \bibnamefont {Wang}},\
  and\ \bibinfo {author} {\bibfnamefont {H.~J.}\ \bibnamefont {Zhao}},\ }\href
  {https://doi.org/10.1007/JHEP09(2022)035} {\bibfield  {journal} {\bibinfo
  {journal} {{J. High Energy Phys.}}\ }\textbf {\bibinfo {volume} {09}},\
  \bibinfo {pages} {035} (\bibinfo {year} {2022})}\BibitemShut {NoStop}%
\bibitem [{\citenamefont {Zhong}\ \emph {et~al.}()\citenamefont {Zhong},
  \citenamefont {Xu},\ and\ \citenamefont {Cheng}}]{Zhong:2024zme}%
  \BibitemOpen
  \bibfield  {author} {\bibinfo {author} {\bibfnamefont {H.}~\bibnamefont
  {Zhong}}, \bibinfo {author} {\bibfnamefont {F.}~\bibnamefont {Xu}},\ and\
  \bibinfo {author} {\bibfnamefont {H.~Y.}\ \bibnamefont {Cheng}},\ }\href@noop
  {} {\ }\Eprint {https://arxiv.org/abs/{2401.15926}} {arXiv:{2401.15926}}
  \BibitemShut {NoStop}%
\bibitem [{\citenamefont {Geng}\ \emph {et~al.}(2018)\citenamefont {Geng},
  \citenamefont {Hsiao}, \citenamefont {Liu},\ and\ \citenamefont
  {Tsai}}]{Geng:2018plk}%
  \BibitemOpen
  \bibfield  {author} {\bibinfo {author} {\bibfnamefont {C.~Q.}\ \bibnamefont
  {Geng}}, \bibinfo {author} {\bibfnamefont {Y.~K.}\ \bibnamefont {Hsiao}},
  \bibinfo {author} {\bibfnamefont {C.~W.}\ \bibnamefont {Liu}},\ and\ \bibinfo
  {author} {\bibfnamefont {T.~H.}\ \bibnamefont {Tsai}},\ }\href
  {https://doi.org/10.1103/PhysRevD.97.073006} {\bibfield  {journal} {\bibinfo
  {journal} {{Phys. Rev. D}}\ }\textbf {\bibinfo {volume} {97}},\ \bibinfo
  {pages} {073006} (\bibinfo {year} {2018})}\BibitemShut {NoStop}%
\bibitem [{\citenamefont {Geng}\ \emph {et~al.}(2019)\citenamefont {Geng},
  \citenamefont {Liu},\ and\ \citenamefont {Tsai}}]{Geng:2019xbo}%
  \BibitemOpen
  \bibfield  {author} {\bibinfo {author} {\bibfnamefont {C.~Q.}\ \bibnamefont
  {Geng}}, \bibinfo {author} {\bibfnamefont {C.~W.}\ \bibnamefont {Liu}},\ and\
  \bibinfo {author} {\bibfnamefont {T.~H.}\ \bibnamefont {Tsai}},\ }\href
  {https://doi.org/10.1016/j.physletb.2019.05.024} {\bibfield  {journal}
  {\bibinfo  {journal} {{Phys. Lett. B}}\ }\textbf {\bibinfo {volume} {794}},\
  \bibinfo {pages} {19} (\bibinfo {year} {2019})}\BibitemShut {NoStop}%
\bibitem [{\citenamefont {Zhong}\ \emph {et~al.}(2023)\citenamefont {Zhong},
  \citenamefont {Xu}, \citenamefont {Wen},\ and\ \citenamefont
  {Gu}}]{Zhong:2022exp}%
  \BibitemOpen
  \bibfield  {author} {\bibinfo {author} {\bibfnamefont {H.}~\bibnamefont
  {Zhong}}, \bibinfo {author} {\bibfnamefont {F.}~\bibnamefont {Xu}}, \bibinfo
  {author} {\bibfnamefont {Q.}~\bibnamefont {Wen}},\ and\ \bibinfo {author}
  {\bibfnamefont {Y.}~\bibnamefont {Gu}},\ }\href
  {https://doi.org/10.1007/JHEP02(2023)235} {\bibfield  {journal} {\bibinfo
  {journal} {{J. High Energy Phys.}}\ }\textbf {\bibinfo {volume} {02}},\
  \bibinfo {pages} {235} (\bibinfo {year} {2023})}\BibitemShut {NoStop}%
\bibitem [{\citenamefont {Xing}\ \emph {et~al.}(2024)\citenamefont {Xing},
  \citenamefont {Shi}, \citenamefont {Sun},\ and\ \citenamefont
  {Xing}}]{Xing:2024nvg}%
  \BibitemOpen
  \bibfield  {author} {\bibinfo {author} {\bibfnamefont {Z.~P.}\ \bibnamefont
  {Xing}}, \bibinfo {author} {\bibfnamefont {Y.~J.}\ \bibnamefont {Shi}},
  \bibinfo {author} {\bibfnamefont {J.}~\bibnamefont {Sun}},\ and\ \bibinfo
  {author} {\bibfnamefont {Y.}~\bibnamefont {Xing}},\ }\href
  {https://doi.org/10.1140/epjc/s10052-024-13389-y} {\bibfield  {journal}
  {\bibinfo  {journal} {{Eur. Phys. J. C}}\ }\textbf {\bibinfo {volume} {84}},\
  \bibinfo {pages} {1014} (\bibinfo {year} {2024})}\BibitemShut {NoStop}%
\bibitem [{\citenamefont {Zou}\ \emph {et~al.}(2020)\citenamefont {Zou},
  \citenamefont {Xu}, \citenamefont {Meng},\ and\ \citenamefont
  {Cheng}}]{Zou:2019kzq}%
  \BibitemOpen
  \bibfield  {author} {\bibinfo {author} {\bibfnamefont {J.}~\bibnamefont
  {Zou}}, \bibinfo {author} {\bibfnamefont {F.}~\bibnamefont {Xu}}, \bibinfo
  {author} {\bibfnamefont {G.}~\bibnamefont {Meng}},\ and\ \bibinfo {author}
  {\bibfnamefont {H.~Y.}\ \bibnamefont {Cheng}},\ }\href
  {https://doi.org/10.1103/PhysRevD.101.014011} {\bibfield  {journal} {\bibinfo
   {journal} {{Phys. Rev. D}}\ }\textbf {\bibinfo {volume} {101}},\ \bibinfo
  {pages} {014011} (\bibinfo {year} {2020})}\BibitemShut {NoStop}%
\bibitem [{\citenamefont {Geng}\ \emph {et~al.}(2024)\citenamefont {Geng},
  \citenamefont {He}, \citenamefont {Jin}, \citenamefont {Liu},\ and\
  \citenamefont {Yang}}]{Geng:2023pkr}%
  \BibitemOpen
  \bibfield  {author} {\bibinfo {author} {\bibfnamefont {C.~Q.}\ \bibnamefont
  {Geng}}, \bibinfo {author} {\bibfnamefont {X.~G.}\ \bibnamefont {He}},
  \bibinfo {author} {\bibfnamefont {X.~N.}\ \bibnamefont {Jin}}, \bibinfo
  {author} {\bibfnamefont {C.~W.}\ \bibnamefont {Liu}},\ and\ \bibinfo {author}
  {\bibfnamefont {C.}~\bibnamefont {Yang}},\ }\href
  {https://doi.org/10.1103/PhysRevD.109.L071302} {\bibfield  {journal}
  {\bibinfo  {journal} {{Phys. Rev. D}}\ }\textbf {\bibinfo {volume} {109}},\
  \bibinfo {pages} {L071302} (\bibinfo {year} {2024})}\BibitemShut {NoStop}%
\bibitem [{\citenamefont {Abashian}\ \emph {et~al.}(2002)\citenamefont
  {Abashian} \emph {et~al.}}]{Belle:2000cnh}%
  \BibitemOpen
  \bibfield  {author} {\bibinfo {author} {\bibfnamefont {A.}~\bibnamefont
  {Abashian}} \emph {et~al.} (\bibinfo {collaboration} {Belle collaboration}),\
  }\href {https://doi.org/10.1016/S0168-9002(01)02013-7} {\bibfield  {journal}
  {\bibinfo  {journal} {Nucl. Instrum. Methods Phys. Res., Sect. A}\ }\textbf
  {\bibinfo {volume} {479}},\ \bibinfo {pages} {117} (\bibinfo {year}
  {2002})}\BibitemShut {NoStop}%
\bibitem [{\citenamefont {Kurokawa}\ and\ \citenamefont
  {Kikutani}(2003)}]{Kurokawa:2001nw}%
  \BibitemOpen
  \bibfield  {author} {\bibinfo {author} {\bibfnamefont {S.}~\bibnamefont
  {Kurokawa}}\ and\ \bibinfo {author} {\bibfnamefont {E.}~\bibnamefont
  {Kikutani}},\ }\href {https://doi.org/10.1016/S0168-9002(02)01771-0}
  {\bibfield  {journal} {\bibinfo  {journal} {Nucl. Instrum. Methods Phys.
  Res., Sect. A}\ }\textbf {\bibinfo {volume} {499}},\ \bibinfo {pages} {1}
  (\bibinfo {year} {2003})}\BibitemShut {NoStop}%
\bibitem [{\citenamefont {Abe}\ \emph {et~al.}(2013)\citenamefont {Abe} \emph
  {et~al.}}]{Abe:2013kxa}%
  \BibitemOpen
  \bibfield  {author} {\bibinfo {author} {\bibfnamefont {T.}~\bibnamefont
  {Abe}} \emph {et~al.},\ }\href {https://doi.org/10.1093/ptep/pts102}
  {\bibfield  {journal} {\bibinfo  {journal} {Prog. Theor. Exp. Phys.}\
  }\textbf {\bibinfo {volume} {2013}},\ \bibinfo {pages} {03A001} (\bibinfo
  {year} {2013})}\BibitemShut {NoStop}%
\bibitem [{\citenamefont {Abe}\ \emph {et~al.}()\citenamefont {Abe} \emph
  {et~al.}}]{Abe:2010gxa}%
  \BibitemOpen
  \bibfield  {author} {\bibinfo {author} {\bibfnamefont {T.}~\bibnamefont
  {Abe}} \emph {et~al.} (\bibinfo {collaboration} {Belle II collaboration}),\
  }\href@noop {} {}\Eprint {https://arxiv.org/abs/1011.0352} {arXiv:1011.0352}
  \BibitemShut {NoStop}%
\bibitem [{\citenamefont {Akai}\ \emph {et~al.}(2018)\citenamefont {Akai},
  \citenamefont {Furukawa},\ and\ \citenamefont {Koiso}}]{Akai:2018mbz}%
  \BibitemOpen
  \bibfield  {author} {\bibinfo {author} {\bibfnamefont {K.}~\bibnamefont
  {Akai}}, \bibinfo {author} {\bibfnamefont {K.}~\bibnamefont {Furukawa}},\
  and\ \bibinfo {author} {\bibfnamefont {H.}~\bibnamefont {Koiso}},\ }\href
  {https://doi.org/10.1016/j.nima.2018.08.017} {\bibfield  {journal} {\bibinfo
  {journal} {Nucl. Instrum. Meth. A}\ }\textbf {\bibinfo {volume} {907}},\
  \bibinfo {pages} {188} (\bibinfo {year} {2018})}\BibitemShut {NoStop}%
\bibitem [{\citenamefont {Adamczyk}\ \emph {et~al.}(2022)\citenamefont
  {Adamczyk} \emph {et~al.}}]{Belle-IISVD:2022upf}%
  \BibitemOpen
  \bibfield  {author} {\bibinfo {author} {\bibfnamefont {K.}~\bibnamefont
  {Adamczyk}} \emph {et~al.} (\bibinfo {collaboration} {Belle II SVD
  collaboration}),\ }\href {https://doi.org/10.1088/1748-0221/17/11/P11042}
  {\bibfield  {journal} {\bibinfo  {journal} {{JINST}}\ }\textbf {\bibinfo
  {volume} {17}},\ \bibinfo {pages} {P11042} (\bibinfo {year}
  {2022})}\BibitemShut {NoStop}%
\bibitem [{\citenamefont {Atmacan}\ \emph {et~al.}(2025)\citenamefont {Atmacan}
  \emph {et~al.}}]{Atmacan:2025jmh}%
  \BibitemOpen
  \bibfield  {author} {\bibinfo {author} {\bibfnamefont {H.}~\bibnamefont
  {Atmacan}} \emph {et~al.},\ }\href
  {https://doi.org/10.1016/j.nima.2025.170627} {\bibfield  {journal} {\bibinfo
  {journal} {Nucl. Instrum. Meth. A}\ }\textbf {\bibinfo {volume} {1080}},\
  \bibinfo {pages} {170627} (\bibinfo {year} {2025})}\BibitemShut {NoStop}%
\bibitem [{\citenamefont {Brodzicka}\ \emph {et~al.}(2012)\citenamefont
  {Brodzicka} \emph {et~al.}}]{Brodzicka:2012jm}%
  \BibitemOpen
  \bibfield  {author} {\bibinfo {author} {\bibfnamefont {J.}~\bibnamefont
  {Brodzicka}} \emph {et~al.} (\bibinfo {collaboration} {Belle
  collaboration}),\ }\href {https://doi.org/10.1093/ptep/pts072} {\bibfield
  {journal} {\bibinfo  {journal} {PTEP}\ }\textbf {\bibinfo {volume} {2012}},\
  \bibinfo {pages} {04D001} (\bibinfo {year} {2012})}\BibitemShut {NoStop}%
\bibitem [{\citenamefont {Adachi}\ \emph
  {et~al.}(2025{\natexlab{b}})\citenamefont {Adachi} \emph
  {et~al.}}]{Belle-II:2024vuc}%
  \BibitemOpen
  \bibfield  {author} {\bibinfo {author} {\bibfnamefont {I.}~\bibnamefont
  {Adachi}} \emph {et~al.} (\bibinfo {collaboration} {Belle II
  collaboration}),\ }\href {https://doi.org/10.1088/1674-1137/ad806c}
  {\bibfield  {journal} {\bibinfo  {journal} {Chin. Phys. C}\ }\textbf
  {\bibinfo {volume} {49}},\ \bibinfo {pages} {013001} (\bibinfo {year}
  {2025}{\natexlab{b}})}\BibitemShut {NoStop}%
\bibitem [{\citenamefont {Lange}(2001)}]{Lange:2001uf}%
  \BibitemOpen
  \bibfield  {author} {\bibinfo {author} {\bibfnamefont {D.~J.}\ \bibnamefont
  {Lange}},\ }\href {https://doi.org/10.1016/S0168-9002(01)00089-4} {\bibfield
  {journal} {\bibinfo  {journal} {Nucl. Instrum. Meth. A}\ }\textbf {\bibinfo
  {volume} {462}},\ \bibinfo {pages} {152} (\bibinfo {year}
  {2001})}\BibitemShut {NoStop}%
\bibitem [{\citenamefont {Sjostrand}\ \emph {et~al.}(2001)\citenamefont
  {Sjostrand}, \citenamefont {Eden}, \citenamefont {Friberg}, \citenamefont
  {Lonnblad}, \citenamefont {Miu}, \citenamefont {Mrenna},\ and\ \citenamefont
  {Norrbin}}]{Sjostrand:2000wi}%
  \BibitemOpen
  \bibfield  {author} {\bibinfo {author} {\bibfnamefont {T.}~\bibnamefont
  {Sjostrand}}, \bibinfo {author} {\bibfnamefont {P.}~\bibnamefont {Eden}},
  \bibinfo {author} {\bibfnamefont {C.}~\bibnamefont {Friberg}}, \bibinfo
  {author} {\bibfnamefont {L.}~\bibnamefont {Lonnblad}}, \bibinfo {author}
  {\bibfnamefont {G.}~\bibnamefont {Miu}}, \bibinfo {author} {\bibfnamefont
  {S.}~\bibnamefont {Mrenna}},\ and\ \bibinfo {author} {\bibfnamefont
  {E.}~\bibnamefont {Norrbin}},\ }\href
  {https://doi.org/10.1016/S0010-4655(00)00236-8} {\bibfield  {journal}
  {\bibinfo  {journal} {Comput. Phys. Commun.}\ }\textbf {\bibinfo {volume}
  {135}},\ \bibinfo {pages} {238} (\bibinfo {year} {2001})}\BibitemShut
  {NoStop}%
\bibitem [{\citenamefont {Jadach}\ \emph {et~al.}(2000)\citenamefont {Jadach},
  \citenamefont {Ward},\ and\ \citenamefont {W\c{a}s}}]{Jadach:1999vf}%
  \BibitemOpen
  \bibfield  {author} {\bibinfo {author} {\bibfnamefont {S.}~\bibnamefont
  {Jadach}}, \bibinfo {author} {\bibfnamefont {B.~F.~L.}\ \bibnamefont
  {Ward}},\ and\ \bibinfo {author} {\bibfnamefont {Z.}~\bibnamefont
  {W\c{a}s}},\ }\href {https://doi.org/10.1016/S0010-4655(00)00048-5}
  {\bibfield  {journal} {\bibinfo  {journal} {Comput. Phys. Commun.}\ }\textbf
  {\bibinfo {volume} {130}},\ \bibinfo {pages} {260} (\bibinfo {year}
  {2000})}\BibitemShut {NoStop}%
\bibitem [{\citenamefont {Sj\"{o}strand}\ \emph {et~al.}(2015)\citenamefont
  {Sj\"{o}strand}, \citenamefont {Ask}, \citenamefont {Christiansen},
  \citenamefont {Corke}, \citenamefont {Desai}, \citenamefont {Ilten},
  \citenamefont {Mrenna}, \citenamefont {Prestel}, \citenamefont {Rasmussen},\
  and\ \citenamefont {Skands}}]{Sjostrand:2014zea}%
  \BibitemOpen
  \bibfield  {author} {\bibinfo {author} {\bibfnamefont {T.}~\bibnamefont
  {Sj\"{o}strand}}, \bibinfo {author} {\bibfnamefont {S.}~\bibnamefont {Ask}},
  \bibinfo {author} {\bibfnamefont {J.~R.}\ \bibnamefont {Christiansen}},
  \bibinfo {author} {\bibfnamefont {R.}~\bibnamefont {Corke}}, \bibinfo
  {author} {\bibfnamefont {N.}~\bibnamefont {Desai}}, \bibinfo {author}
  {\bibfnamefont {P.}~\bibnamefont {Ilten}}, \bibinfo {author} {\bibfnamefont
  {S.}~\bibnamefont {Mrenna}}, \bibinfo {author} {\bibfnamefont
  {S.}~\bibnamefont {Prestel}}, \bibinfo {author} {\bibfnamefont {C.~O.}\
  \bibnamefont {Rasmussen}},\ and\ \bibinfo {author} {\bibfnamefont {P.~Z.}\
  \bibnamefont {Skands}},\ }\href {https://doi.org/10.1016/j.cpc.2015.01.024}
  {\bibfield  {journal} {\bibinfo  {journal} {Comput. Phys. Commun.}\ }\textbf
  {\bibinfo {volume} {191}},\ \bibinfo {pages} {159} (\bibinfo {year}
  {2015})}\BibitemShut {NoStop}%
\bibitem [{\citenamefont {Chan}\ \emph {et~al.}(2001)\citenamefont {Chan} \emph
  {et~al.}}]{CLEO:2000lsg}%
  \BibitemOpen
  \bibfield  {author} {\bibinfo {author} {\bibfnamefont {S.}~\bibnamefont
  {Chan}} \emph {et~al.} (\bibinfo {collaboration} {CLEO collaboration}),\ }\href
  {https://doi.org/10.1103/PhysRevD.63.111102} {\bibfield  {journal} {\bibinfo
  {journal} {Phys. Rev. D}\ }\textbf {\bibinfo {volume} {63}},\ \bibinfo
  {pages} {111102} (\bibinfo {year} {2001})}\BibitemShut {NoStop}%
\bibitem [{\citenamefont {Barberio}\ \emph {et~al.}(1991)\citenamefont
  {Barberio}, \citenamefont {van Eijk},\ and\ \citenamefont
  {W\c{a}s}}]{Barberio:1990ms}%
  \BibitemOpen
  \bibfield  {author} {\bibinfo {author} {\bibfnamefont {E.}~\bibnamefont
  {Barberio}}, \bibinfo {author} {\bibfnamefont {B.}~\bibnamefont {van Eijk}},\
  and\ \bibinfo {author} {\bibfnamefont {Z.}~\bibnamefont {W\c{a}s}},\ }\href
  {https://doi.org/10.1016/0010-4655(91)90012-A} {\bibfield  {journal}
  {\bibinfo  {journal} {Comput. Phys. Commun.}\ }\textbf {\bibinfo {volume}
  {66}},\ \bibinfo {pages} {115} (\bibinfo {year} {1991})}\BibitemShut
  {NoStop}%
\bibitem [{\citenamefont {Brun}\ \emph {et~al.}(1987)\citenamefont {Brun},
  \citenamefont {Bruyant}, \citenamefont {Maire}, \citenamefont {McPherson},\
  and\ \citenamefont {Zanarini}}]{Brun:1987ma}%
  \BibitemOpen
  \bibfield  {author} {\bibinfo {author} {\bibfnamefont {R.}~\bibnamefont
  {Brun}}, \bibinfo {author} {\bibfnamefont {F.}~\bibnamefont {Bruyant}},
  \bibinfo {author} {\bibfnamefont {M.}~\bibnamefont {Maire}}, \bibinfo
  {author} {\bibfnamefont {A.~C.}\ \bibnamefont {McPherson}},\ and\ \bibinfo
  {author} {\bibfnamefont {P.}~\bibnamefont {Zanarini}},\ }\href@noop {} {\
  (\bibinfo {year} {1987})},\ \Eprint {https://arxiv.org/abs/{CERN-DD-EE-84-1}}
  {{CERN-DD-EE-84-1}} \BibitemShut {NoStop}%
\bibitem [{\citenamefont {Agostinelli}\ \emph {et~al.}(2003)\citenamefont
  {Agostinelli} \emph {et~al.}}]{Agostinelli:2002hh}%
  \BibitemOpen
  \bibfield  {author} {\bibinfo {author} {\bibfnamefont {S.}~\bibnamefont
  {Agostinelli}} \emph {et~al.} (\bibinfo {collaboration} {GEANT4
  collaboration}),\ }\href {https://doi.org/10.1016/S0168-9002(03)01368-8}
  {\bibfield  {journal} {\bibinfo  {journal} {Nucl. Instrum. Meth. A}\ }\textbf
  {\bibinfo {volume} {506}},\ \bibinfo {pages} {250} (\bibinfo {year}
  {2003})}\BibitemShut {NoStop}%
\bibitem [{\citenamefont {Zhou}\ \emph {et~al.}(2021)\citenamefont {Zhou},
  \citenamefont {Du}, \citenamefont {Li},\ and\ \citenamefont
  {Shen}}]{Zhou:2020ksj}%
  \BibitemOpen
  \bibfield  {author} {\bibinfo {author} {\bibfnamefont {X.}~\bibnamefont
  {Zhou}}, \bibinfo {author} {\bibfnamefont {S.}~\bibnamefont {Du}}, \bibinfo
  {author} {\bibfnamefont {G.}~\bibnamefont {Li}},\ and\ \bibinfo {author}
  {\bibfnamefont {C.}~\bibnamefont {Shen}},\ }\href
  {https://doi.org/10.1016/j.cpc.2020.107540} {\bibfield  {journal} {\bibinfo
  {journal} {{Comput. Phys. Commun.}}\ }\textbf {\bibinfo {volume} {258}},\
  \bibinfo {pages} {107540} (\bibinfo {year} {2021})}\BibitemShut {NoStop}%
\bibitem [{\citenamefont {Gelb}\ \emph {et~al.}(2018)\citenamefont {Gelb} \emph
  {et~al.}}]{Gelb:2018agf}%
  \BibitemOpen
  \bibfield  {author} {\bibinfo {author} {\bibfnamefont {M.}~\bibnamefont
  {Gelb}} \emph {et~al.},\ }\href {https://doi.org/10.1007/s41781-018-0016-x}
  {\bibfield  {journal} {\bibinfo  {journal} {Comput. Softw. Big Sci.}\
  }\textbf {\bibinfo {volume} {2}},\ \bibinfo {pages} {9} (\bibinfo {year}
  {2018})}\BibitemShut {NoStop}%
\bibitem [{\citenamefont {Kuhr}\ \emph {et~al.}(2019)\citenamefont {Kuhr},
  \citenamefont {Pulvermacher}, \citenamefont {Ritter}, \citenamefont {Hauth},\
  and\ \citenamefont {Braun}}]{Kuhr:2018lps}%
  \BibitemOpen
  \bibfield  {author} {\bibinfo {author} {\bibfnamefont {T.}~\bibnamefont
  {Kuhr}}, \bibinfo {author} {\bibfnamefont {C.}~\bibnamefont {Pulvermacher}},
  \bibinfo {author} {\bibfnamefont {M.}~\bibnamefont {Ritter}}, \bibinfo
  {author} {\bibfnamefont {T.}~\bibnamefont {Hauth}},\ and\ \bibinfo {author}
  {\bibfnamefont {N.}~\bibnamefont {Braun}} (\bibinfo {collaboration} {Belle II
  Framework Software Group}),\ }\href
  {https://doi.org/10.1007/s41781-018-0017-9} {\bibfield  {journal} {\bibinfo
  {journal} {Comput. Softw. Big Sci.}\ }\textbf {\bibinfo {volume} {3}},\
  \bibinfo {pages} {1} (\bibinfo {year} {2019})}\BibitemShut {NoStop}%
\bibitem [{\citenamefont {Punzi}(2003{\natexlab{a}})}]{Punzi:2003bu}%
  \BibitemOpen
  \bibfield  {author} {\bibinfo {author} {\bibfnamefont {G.}~\bibnamefont
  {Punzi}},\ }\href@noop {} {\bibfield  {journal} {\bibinfo  {journal} {eConf}\
  }\textbf {\bibinfo {volume} {C030908}},\ \bibinfo {pages} {MODT002} (\bibinfo
  {year} {2003}{\natexlab{a}})}\BibitemShut {NoStop}%
\bibitem [{\citenamefont {Punzi}(2003{\natexlab{b}})}]{Punzi:2003wze}%
  \BibitemOpen
  \bibfield  {author} {\bibinfo {author} {\bibfnamefont {G.}~\bibnamefont
  {Punzi}},\ }\href@noop {} {\bibfield  {journal} {\bibinfo  {journal} {eConf}\
  }\textbf {\bibinfo {volume} {C030908}},\ \bibinfo {pages} {WELT002} (\bibinfo
  {year} {2003}{\natexlab{b}})}\BibitemShut {NoStop}%
\bibitem [{\citenamefont {Feichtinger}\ \emph {et~al.}(2022)\citenamefont
  {Feichtinger} \emph {et~al.}}]{Feichtinger:2021uff}%
  \BibitemOpen
  \bibfield  {author} {\bibinfo {author} {\bibfnamefont {P.}~\bibnamefont
  {Feichtinger}} \emph {et~al.},\ }\href
  {https://doi.org/10.1140/epjc/s10052-022-10070-0} {\bibfield  {journal}
  {\bibinfo  {journal} {Eur. Phys. J. C}\ }\textbf {\bibinfo {volume} {82}},\
  \bibinfo {pages} {121} (\bibinfo {year} {2022})}\BibitemShut {NoStop}%
\bibitem [{\citenamefont {Keck}(2017)}]{Keck:2017gsv}%
  \BibitemOpen
  \bibfield  {author} {\bibinfo {author} {\bibfnamefont {T.}~\bibnamefont
  {Keck}},\ }\href {https://doi.org/10.1007/s41781-017-0002-8} {\bibfield
  {journal} {\bibinfo  {journal} {Comput. Softw. Big Sci.}\ }\textbf {\bibinfo
  {volume} {1}},\ \bibinfo {pages} {2} (\bibinfo {year} {2017})}\BibitemShut
  {NoStop}%
\bibitem [{\citenamefont {Cheema}(2024)}]{Cheema:2024iek}%
  \BibitemOpen
  \bibfield  {author} {\bibinfo {author} {\bibfnamefont {P.}~\bibnamefont
  {Cheema}},\ }\href {https://doi.org/10.1051/epjconf/202429509035} {\bibfield
  {journal} {\bibinfo  {journal} {EPJ Web Conf.}\ }\textbf {\bibinfo {volume}
  {295}},\ \bibinfo {pages} {09035} (\bibinfo {year} {2024})}\BibitemShut
  {NoStop}%
\bibitem [{\citenamefont {Krohn}\ \emph {et~al.}(2020)\citenamefont {Krohn}
  \emph {et~al.}}]{Krohn:2019dlq}%
  \BibitemOpen
  \bibfield  {author} {\bibinfo {author} {\bibfnamefont {J.~F.}\ \bibnamefont
  {Krohn}} \emph {et~al.} (\bibinfo {collaboration} {Belle II Analysis Software
  Group}),\ }\href {https://doi.org/10.1016/j.nima.2020.164269} {\bibfield
  {journal} {\bibinfo  {journal} {Nucl. Instrum. Meth. A}\ }\textbf {\bibinfo
  {volume} {976}},\ \bibinfo {pages} {164269} (\bibinfo {year}
  {2020})}\BibitemShut {NoStop}%
\bibitem [{\citenamefont {Wilks}(1938)}]{Wilks:1938dza}%
  \BibitemOpen
  \bibfield  {author} {\bibinfo {author} {\bibfnamefont {S.~S.}\ \bibnamefont
  {Wilks}},\ }\href {https://doi.org/10.1214/aoms/1177732360} {\bibfield
  {journal} {\bibinfo  {journal} {Annals Math. Statist.}\ }\textbf {\bibinfo
  {volume} {9}},\ \bibinfo {pages} {60} (\bibinfo {year} {1938})}\BibitemShut
  {NoStop}%
\bibitem [{\citenamefont {Bertacchi}\ \emph {et~al.}(2021)\citenamefont
  {Bertacchi} \emph {et~al.}}]{BelleIITrackingGroup:2020hpx}%
  \BibitemOpen
  \bibfield  {author} {\bibinfo {author} {\bibfnamefont {V.}~\bibnamefont
  {Bertacchi}} \emph {et~al.} (\bibinfo {collaboration} {Belle II Tracking
  Group}),\ }\href {https://doi.org/10.1016/j.cpc.2020.107610} {\bibfield
  {journal} {\bibinfo  {journal} {Comput. Phys. Commun.}\ }\textbf {\bibinfo
  {volume} {259}},\ \bibinfo {pages} {107610} (\bibinfo {year}
  {2021})}\BibitemShut {NoStop}%
\bibitem [{\citenamefont {Nakano}(2002)}]{Nakano:2002jw}%
  \BibitemOpen
  \bibfield  {author} {\bibinfo {author} {\bibfnamefont {E.}~\bibnamefont
  {Nakano}},\ }\href {https://doi.org/10.1016/S0168-9002(02)01510-3} {\bibfield
   {journal} {\bibinfo  {journal} {Nucl. Instrum. Meth. A}\ }\textbf {\bibinfo
  {volume} {494}},\ \bibinfo {pages} {402} (\bibinfo {year}
  {2002})}\BibitemShut {NoStop}%
\bibitem [{\citenamefont {Adachi}\ \emph
  {et~al.}(2025{\natexlab{c}})\citenamefont {Adachi} \emph
  {et~al.}}]{Belle-II:2025tpe}%
  \BibitemOpen
  \bibfield  {author} {\bibinfo {author} {\bibfnamefont {I.}~\bibnamefont
  {Adachi}} \emph {et~al.} (\bibinfo {collaboration} {Belle II
  collaboration}),\ }\href {https://doi.org/10.1140/epjc/s10052-025-14627-7}
  {\bibfield  {journal} {\bibinfo  {journal} {Eur. Phys. J. C}\ }\textbf
  {\bibinfo {volume} {85}},\ \bibinfo {pages} {1237} (\bibinfo {year}
  {2025}{\natexlab{c}})}\BibitemShut {NoStop}%
\end{thebibliography}
\end{document}